# Linear Analysis of Boundary-Layer Instabilities on a Finned-Cone at Mach 6

Daniel B. Araya[1], Neal P. Bitter[2], Bradley M. Wheaton[3]
*Johns Hopkins University Applied Physics Laboratory, Laurel, MD 20723*

Omar Kamal[4], Tim Colonius[5]
*California Institute of Technology, Pasadena, CA 91125*

Anthony Knutson[6], Heath Johnson[7], Joseph Nichols[8], Graham V. Candler[9]
*University of Minnesota, Minneapolis, MN 55455*

Vincenzo Russo[10]
*University of Kentucky, Lexington, KY 40506*

Christoph Brehm[11]
*University of Maryland, College Park, MD 20742*

**Boundary-layer instabilities for a finned cone at Mach = 6, Re = $8.4 \times 10^6$ [$m^{-1}$], and zero incidence angle are examined using linear stability methods of varying fidelity and maturity, following earlier analysis presented in [1]. The geometry and laminar flow conditions correspond to experiments conducted at the Boeing Air Force Mach 6 Quiet Tunnel (BAM6QT) at Purdue University. Where possible, a common mean flow is utilized among the stability computations, and comparisons are made along the acreage of the cone where transition is first observed in the experiment. Stability results utilizing Linear Stability Theory (LST), planar Parabolized Stability Equations (planar-PSE), One-Way Navier Stokes (OWNS), forced direct numerical simulation (DNS), and Adaptive Mesh Refinement Wavepacket Tracking (AMR-WPT) are presented. A dominant three-dimensional vortex instability occurring at $\approx$ 250 kHz is identified that correlates well with experimental measurements of transition onset. With the exception of LST, all of the higher-fidelity linear methods considered in this work were consistent in predicting the initial growth and general structure of the vortex instability as it evolved downstream. Some of the challenges, opportunities, and development needs of the stability methods considered are discussed.**

[1]Senior Aerospace Engineer, AIAA Senior Member
[2]Senior Aerospace Engineer, AIAA Senior Member
[3]Senior Aerospace Engineer, AIAA Associate Fellow
[4]Graduate Research Assistant, AIAA Member
[5]Professor, Associate Fellow AIAA
[6]Senior Research Associate, AIAA Member
[7]Senior Research Associate, AIAA Senior Member
[8]Associate Professor, AIAA Senior Member
[9]Professor, AIAA Fellow.
[10]Graduate Research Assistant, AIAA Member
[11]Assistant Professor, AIAA Senior Member

# I. Introduction

Practical computations of hypersonic boundary-layer stability were pioneered by Mack [2] and remain today an essential part of transition analysis for the design of hypersonic vehicles; see [3] and [4] for a historical perspective. The accurate predictions of classical linear stability theory for simple axisymmetric shapes such as a sharp cone are well-documented (see e.g., [5]). However, there are limits to applying the theory for complex shapes. Three-dimensional flows complicate the stability analysis with the presence of multiple instability mechanisms and possible nonlinear interactions [6]. A number of modern numerical tools exist today for more comprehensive characterization of linear amplification mechanisms in these complex flows, and the goal of this paper is to demonstrate the application of a few of these methods to a complex flow field of practical relevance.

From an engineering perspective, predicting the large changes in heat transfer and aerodynamic performance associated with a hypersonic boundary layer transitioning from laminar to turbulent flow remains a fundamental challenge in vehicle design. The transition process is known to be sensitive to initial and boundary conditions and this creates large uncertainty in aerothermal predictions that engineers must manage with large margins. The state of practice for predicting the onset of transition on hypersonic vehicles is to use simplistic correlations, such as ones based on historical flight test data from sphere-cone geometries (see e.g., [7]). Although sometimes useful, correlations lack any underlying physical model of the flow and thus also lack general predictive power.

Boundary-layer Linear Stability Theory (LST) utilizing the so-called parallel-flow approximation has been used for several decades to examine the growth of small modal, i.e., wave-like, disturbances to a laminar boundary layer [2]. The extension of LST to Parabolized Stability Equations (PSE) can account for weakly non-parallel and nonlinear effects and has been in use since the early 1990's [8–10]. Today, numerical LST and PSE solvers can be used routinely to determine the growth of modal disturbances and arrive at boundary-layer transition predictions based on an N-factor, a semi-empirical method first proposed by [11, 12]. The larger the N-factor, the greater the exponential growth of the disturbance and typically a threshold value (e.g., N=10) is used to estimate the onset of transition, either correlated from experiments or based on past experience. Although widely used, constant N-factor transition predictions are known to be inaccurate (see e.g. [13]) because they can be mechanism and condition dependent. Furthermore, the underlying assumptions of classical LST and PSE limit their application to flows with strong variation in just one dimension and require an ad hoc choice of disturbance paths, so they must be used with caution when applied to vehicles with complex geometric features, such as fins.

Over the last two decades, a number of methods for predicting the initial growth of boundary-layer instabilities on complex shapes in hypersonic flow have been developed. These methods include spatial BiGlobal analysis [14, 15], plane-marching PSE [16, 17], One-Way Navier Stokes (OWNS) [18], linearized Direct Numerical Simulation (LDNS) [19], forced DNS coupled with sparsity-promoting Dynamic Mode Decomposition [20], Wavepacket Tracking with Adaptive Mesh Refinement (AMR-WPT) [21], and Input/Output (I/O) analysis [22]. All of these methods are linear and therefore some transition mechanisms, such as the growth of secondary instabilities, are not captured. The justification for focusing on linear methods is rooted in the fact that atmospheric flight conditions are known to be nominally a low-disturbance environment. This fact, along with the ability to accommodate complex three-dimensional flow fields, makes these methods attractive for application to transition prediction on real vehicles.

The complex geometry considered in this work is a 7° half-angle cone with 1 mm nose radius and a fin with 0.125 in leading edge radius and 75° sweep angle, as shown in figure 1. This geometry was built for boundary-layer transition experiments conducted in the Mach 6 quiet tunnel at Purdue University and further details are described in [23–26]. The finned cone is of practical interest since the fin represents a basic means to control a high-speed flight vehicle. The addition of the fin introduces a shock-boundary layer interaction on the cone surface due to the fin shock as well as a three-dimensional streamwise-aligned corner flow from the fin-cone junction. Both of these flow features complicate the computational analysis due to the spanwise distortion of the laminar flow. Several groups have made progress over the last few years in studying the surface heating and boundary-layer stability characteristics of various fin-cone configurations tested at Purdue [20, 27–33]. From this prior analysis, it was clear that streamwise-aligned laminar vortices played an important role in the transition process along the cone, but the question of how to interpret the results of different linear stability analysis methods when applied to this complex flow field remained unclear.

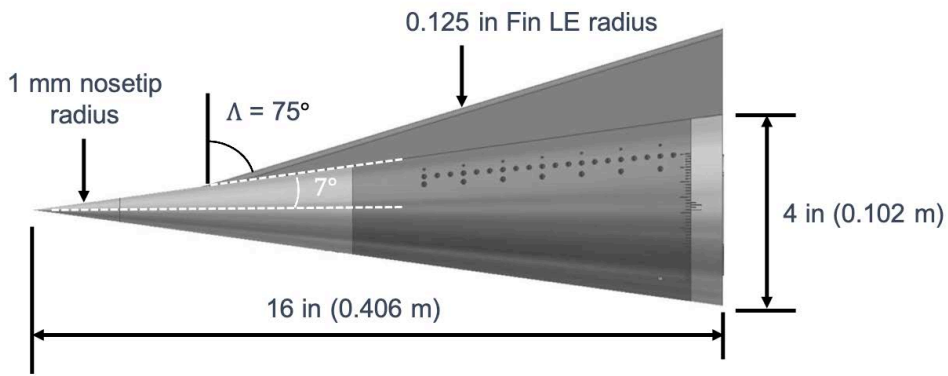

**Fig. 1  Schematic of Purdue finned-cone wind tunnel model utillized. Adapted with permission from [24].**

This question was first addressed in an earlier publication by the authors [1] and is more comprehensively examined in the current paper, with the overall goal being to provide a better understanding of the use of linear stability methods for analyzing complex hypersonic boundary-layer flows. Specifically, we apply several of the aforementioned linear stability analysis methods to the laminar flow over the acreage of a finned cone at Mach = 6, $Re = 8.4 \times 10^6$ $[m^{-1}]$, and zero incidence angle. The results presented in this work are for a specific flow field that may be well-suited to one or more of the linear methods or codes utilized. This is instructive from a practical perspective but it is not intended to be comprehensive or to promote one method or code above the rest, since each are at different development stages and have advantages for use depending on the flow considered and objectives of the analysis. The remainder of the paper is organized as follows: section II presents a summary of relevant linear stability theory; section III describes the computational methods utilized; section IV defines the common laminar mean flow and necessary adaptations for each method; section V presents the stability results; finally, section VI discusses the challenges, opportunities, and future development needs of the methods for this and other complex flow problems.

## II. Linear Stability Analysis Methods

The governing equations for the fluid system under consideration are the nonlinear compressible Navier-Stokes equations, which in dimensional form can be expressed as the following:

$$\frac{\partial \rho}{\partial t} + \nabla \cdot (\rho \mathbf{u}) = 0 \tag{1a}$$

$$\rho \frac{\partial \mathbf{u}}{\partial t} + \rho \mathbf{u} \cdot \nabla \mathbf{u} + \nabla p = \nabla \cdot \boldsymbol{\tau} \tag{1b}$$

$$\rho C_p \left( \frac{\partial T}{\partial t} + \mathbf{u} \cdot \nabla T \right) + p \nabla \cdot \mathbf{u} = \nabla \cdot (k \nabla T) + \mathbf{u} : \boldsymbol{\tau} \tag{1c}$$

$$p = \rho \mathcal{R} T \tag{1d}$$

Here $\rho$ is the density, $p$ the pressure, $T$ the temperature, $C_p$ the specific heat, $\mathcal{R}$ the gas constant, $k$ the thermal conductivity, and $\mathbf{u}$ the velocity vector with Cartesian components $(u, v, w)$ in the $(x, y, z)$ directions. The viscous stress tensor is denoted $\boldsymbol{\tau}$, with components

$$\tau_{ij} = \mu \left( \frac{\partial u_i}{\partial x_j} + \frac{\partial u_j}{\partial x_i} \right) + \lambda \frac{\partial u_k}{\partial x_k} \delta_{ij} \tag{2}$$

In this equation, $\mu$ is the shear viscosity, which is calculated using Sutherland's model, and $\lambda$ is the second viscosity. Its value is determined by applying Stokes' hypothesis of zero bulk viscosity, which implies $\lambda = -2/3\mu$. The symbol $\delta_{ij}$ is the Kroneker delta. In the equations of energy and state (1c-1d), the gas is assumed to behave as a calorically perfect ideal gas, which is appropriate for the low enthalpy wind tunnel conditions under consideration.

With the governing equations defined, the general procedure for a linear analysis of boundary layer instabilities is the following: (1) define a basic state, i.e., the laminar mean flow; (2) superpose a small disturbance onto the basic state and substitute into the governing equations; (3) linearize the equations; (4) impose an ansatz for the spatial form of the disturbance; (5) solve the linearized equations to determine the evolution of the small disturbance; (6) characterize the resulting amplification or attenuation of the disturbance. It is at either step (3) or (4) that the linear methods diverge based on their underlying assumptions. This is summarized in table 1 for the methods considered in this paper, where the instantaneous vector of fluid variables $\mathbf{q} = [\rho, \mathbf{u}, T]^T$ is decomposed in the usual way into steady ($\bar{\mathbf{q}}$) and unsteady ($\mathbf{q}'$) parts, such that $\mathbf{q} = \bar{\mathbf{q}} + \mathbf{q}'$. Note that the partial derivatives in the table refer to a generalized body-fixed coordinate

Table 1  Linear stability methods considered in this paper.

| Linear Method | Mean Flow Variation | Basic State | Shape Function | Phase Function |
|---|---|---|---|---|
| LST | $\partial_\xi \bar{\mathbf{q}} = \partial_\zeta \bar{\mathbf{q}} = 0$ | $\bar{\mathbf{q}}(\eta)$ | $\hat{\mathbf{q}}(\eta)$ | $\Theta = \exp[i(\alpha\xi + \beta\zeta - \omega t)]$ |
| PSE | $\partial_\xi \bar{\mathbf{q}} << \partial_\eta \bar{\mathbf{q}}; \partial_\zeta \bar{\mathbf{q}} = 0$ | $\bar{\mathbf{q}}(\xi,\eta)$ | $\hat{\mathbf{q}}(\xi,\eta)$ | $\Theta = \exp[i(\int \alpha(\xi')d\xi' + \beta\zeta - \omega t)]$ |
| BiGlobal | $\partial_\xi \bar{\mathbf{q}} = 0$ | $\bar{\mathbf{q}}(\eta,\zeta)$ | $\hat{\mathbf{q}}(\eta,\zeta)$ | $\Theta = \exp[i(\alpha\xi - \omega t)]$ |
| Planar PSE | $\partial_\xi \bar{\mathbf{q}} << [\partial_\eta \bar{\mathbf{q}}, \partial_\zeta \bar{\mathbf{q}}]$ | $\bar{\mathbf{q}}(\xi,\eta,\zeta)$ | $\hat{\mathbf{q}}(\xi,\eta,\zeta)$ | $\Theta = \exp[i(\int \alpha(\xi')d\xi' - \omega t)]$ |
| OWNS** | $\partial_\xi \bar{\mathbf{q}} << [\partial_\eta \bar{\mathbf{q}}, \partial_\zeta \bar{\mathbf{q}}]$ | $\bar{\mathbf{q}}(\xi,\eta,\zeta)$ | arbitrary | arbitrary |
| AMR-WPT | arbitrary | $\bar{\mathbf{q}}(\xi,\eta,\zeta)$ | arbitrary | arbitrary |
| Forced DNS | arbitrary | $\bar{\mathbf{q}}(\xi,\eta,\zeta)$ | arbitrary | arbitrary |

**For OWNS, the weakly parallel approximation is only invoked to parabolize the governing equations and is not otherwise imposed on the ansatz.

system, where $(\xi, \eta, \zeta)$ correspond to the streamwise, wall-normal, and spanwise directions, respectively. It is also worth mentioning that Input/Output (I/O) analysis was initially considered as another linear stability method for comparison, due to its applicability to analyzing the stability of arbitrary mean flows, as described in [1]. The interested reader is referred to the reference for a detailed description of the I/O method, including its benefits and associated challenges for the specific numerical implementation that was evaluated.

For methods that consider only eigensolutions to the linearized system of equations, so-called modal methods, the ansatz for the disturbance is of the form $\mathbf{q}' = [\rho', \mathbf{u}', T']^T = \hat{\mathbf{q}}\Theta$, where $\hat{\mathbf{q}}$ is the amplitude, or shape function, and $\Theta$ is the phase function for a wave-like disturbance. For nonmodal methods, which includes OWNS, forced DNS, and AMR-WPT, there is no formal ansatz and the disturbance can be represented by a linear superposition of arbitrary functions in space and time. These arbitrary disturbance functions are not typically referred to as "shape" and "phase" functions in the literature, but they are listed in the table as such to provide perspective with the modal methods.

Regardless of the method, the process of linearizing equations 1a-1c is the same, with the first step being to substitute $\mathbf{q} = \bar{\mathbf{q}} + \mathbf{q}'$ into these equations. Next, the mean flow terms that satisfy the governing equations are subtracted out and terms that are $O(\mathbf{q}'^2)$ are neglected. Details of this derivation can be found in many references, e.g.,[3]. After linearizing the Navier-Stokes equations (c.f. equation set 1) about the steady base flow and transforming them into the generalized coordinates, the linearized equations can be written in the form:

$$\frac{\partial \mathbf{q}'}{\partial t} + \mathbf{A_o}\mathbf{q}' + \mathbf{A}_i \frac{\partial \mathbf{q}'}{\partial x^i} + \mathbf{A}_{ij} \frac{\partial^2 \mathbf{q}'}{\partial x^i \partial x^j} \tag{3}$$

Here $\mathbf{A}_o$, $\mathbf{A}_i$, and $\mathbf{A}_{ij}$ are $5 \times 5$ matrices, $x^i$ is the vector of curvilinear coordinates $(\xi, \eta, \zeta)$, and $\mathbf{q}'$ is the vector of disturbance variables:

$$\mathbf{q}' = \left(\rho', u'_c, v'_c, w'_c, T'\right)^T \tag{4}$$

where the accent mark $'$ is used to denote fluctuating quantities and subscript 'c' indicates that the velocity vector has been transformed into generalized coordinates. As mentioned in the introduction, the linear stability methods considered in this paper are not all-inclusive and are also each at different stages of maturity. To emphasize this point, table 2 lists the methods considered along with an estimate of the number of years that each method has been developed specifically for the stability analysis of hypersonic boundary-layer flows based on available publications. The key differences among the linear methods are in the assumptions made about the baseflow and disturbance quantities. It is up to the analyst to consider the features of the laminar baseflow and pick the form of the small disturbances given in table 1 that are appropriate for the flow under investigation, which are reviewed next.

Table 2  Estimated maturity of linear stability methods considered in this paper.

| Linear Method | Approximate Maturity of Method in Number of Years Since Referenced Publication | Early Publication of Method Applied to Hypersonic Boundary Layers |
|---|---|---|
| LST | 57 | [2] |
| PSE | 29 | [8] |
| Forced DNS | 21 | [34] |
| BiGlobal | 20 | [14] |
| Planar PSE | 11 | [16] |
| AMR-WPT | 5 | [35] |
| OWNS | 2 | [18] |

## A. Flows With Strong Variation in 1D Only

### 1. Linear Stability Theory (LST)

For flows that have strong variation in only one-dimension, the parallel flow approximation can be made to simplify the linearized equations. This approximation implies that the mean flow only varies in the wall-normal direction, i.e. $\bar{\mathbf{q}}(\eta)$ and in classic LST the disturbances take the form of normal modes with constant wavenumber and frequency, i.e.,

$$\mathbf{q}'(\eta, t) = \hat{\mathbf{q}}(\eta) \exp\left[i\left(\alpha\xi + \beta\zeta - \omega t\right)\right]. \tag{5}$$

Here $\hat{\mathbf{q}}(\eta)$ is the amplitude or shape function and $(\alpha, \beta, \omega)$ are the streamwise wavenumber, spanwise wavenumber, and temporal frequency for the disturbance, respectively, which are all constants that can be either real or complex. The typical spatial stability problem for boundary layer flows involves fixing $\beta$ and $\omega$ and solving an eigenvalue problem where the complex eigenvalues are $\alpha = \alpha_r + i\alpha_i$ and each corresponding eigenfunction is the shape function. Disturbances with $\alpha_i < 0$ are unstable and grow and have a corresponding streamwise wavenumber equal to $\alpha_r$.

### 2. Linear Parabolized Stability Equations (PSE)

For flows with strong variation in the wall-normal direction but that also vary slowly in the streamwise direction, the parallel flow approximation can be relaxed. Specifically, the disturbance ansatz is adjusted to allow slow variation in the streamwise wavenumber and has the form

$$\mathbf{q}'(\xi, \eta, t) = \hat{\mathbf{q}}(\xi, \eta) \exp\left[i\left(\int_{\xi_0}^{\xi} \alpha(\xi')d\xi' + \beta\zeta - \omega t\right)\right], \tag{6}$$

where $\xi$ is the slow-varying direction. This modification allows for some weakly-nonparallel baseflow effects to be modeled, while still remaining economical from a computational perspective since the modified governing equations for the amplitude function ($\hat{\mathbf{q}}(\xi, \eta)$) become parabolic. The latter point also gives rise to the name for this form of the equations, which are the Parabolized Stability Equations (PSE).

In equation 6, an ambiguity exists where changes in the disturbance amplitude along the slow varying direction ($\xi$) can be included in either the shape function ($\hat{\mathbf{q}}$) or the phase function ($\Theta$) and must be resolved by a suitable constraint. Further details on the linear PSE formulation and extension to include weakly nonlinear effects for compressible flows are found in [8, 10]. The next section will also discuss how the same PSE method can be extended to flows with strong variation in two dimensions.

**B. Flows With Strong Variation in 2D Only**

*1. BiGlobal and Planar PSE*

Another two types of eigenmode analysis that are applicable to flows with strong mean flow variation in two directions are spatial BiGlobal and planar PSE analysis, with the latter sometimes referred to as PSE-3D in the literature [16]. These methods are directly analogous to LST and PSE of the previous subsection with the extension being that the shape function is now free to vary in a plane rather than just the wall-normal coordinate. For planar PSE analysis, the disturbance vector $\mathbf{q}'$ is the same as in equation 6 except for a change to the shape function to accommodate spanwise ($\zeta$) variation as follows:

$$\mathbf{q}'(\xi, \eta, \zeta, t) = \hat{\mathbf{q}}(\xi, \eta, \zeta) \exp\left(-i\int_{\xi_o}^{\xi} \alpha(\xi')d\xi' - i\omega t\right) \tag{7}$$

where $\hat{\mathbf{q}}$ is the shape function that is slowly varying in the $\xi$ direction and $\alpha$ is a streamwise wavenumber that captures rapid streamwise variation of the disturbance. Just as in classical, line-marching PSE, the decomposition of the disturbance into a wave part $\alpha$ and a shape function $\hat{\mathbf{q}}$ in equation 7 is ambiguous and must be resolved by supplementing a suitable norm to close the formulation.

By substituting equation 7 into the linearized Navier-Stokes equations (3) and dropping second derivatives with respect to $\xi$, one arrives at a linear system of equations of the form:

$$\mathbf{L}_\xi \frac{\partial \hat{\mathbf{q}}}{\partial \xi} + \mathbf{L}_o \hat{\mathbf{q}} = 0, \tag{8}$$

where $\mathbf{L}_\xi$ and $\mathbf{L}_o$ are matrix operators. This is the governing system of equations for planar PSE. Given a suitable initial condition $\hat{\mathbf{q}}_o$ at some initial location $\xi_o$, the three dimensional PSE solution $\hat{\mathbf{q}}(\xi, \eta, \zeta)$ is computed by discretizing and marching equation 8 in the $\xi$ direction using appropriate numerical methods.

If the streamwise derivatives and terms that are quadratic in the wavenumber $\alpha$ are eliminated from equation 8, the system of equations reduces to a linear eigenvalue problem of the form:

$$\mathbf{L}\hat{\mathbf{q}} = \alpha \mathbf{R}\hat{\mathbf{q}} \tag{9}$$

where $\mathbf{L}$ and $\mathbf{R}$ are square matrices of dimension $5 \times N_\eta \times N_\zeta$ and $N_\eta$ and $N_\zeta$ are the number of grid points in the wall-normal and spanwise directions, respectively. This is the spatial BiGlobal eigenvalue problem and solving the system can be done at any streamwise ($\xi$) station in the flow, making it a locally two-dimensional analysis of instabilities. This BiGlobal stability analysis can also be used to provide the initial condition for the planar PSE march as discussed above.

*2. One-Way Navier-Stokes (OWNS)*

OWNS is a streamwise-marching technique that employs a rigorous parabolization method to generate well-posed, one-way approximations and efficiently removes modes with upstream group velocity using recursive filters that were originally developed for non-reflecting boundary conditions. Although more computationally costly than PSE, OWNS alleviates several of PSE's limitations. Instead of formally deriving a one-way operator, PSE achieves a stable spatial march via numerical damping that affects a subset of the downstream-propagating modes in addition to the targeted upstream modes [36]. In either form, the associated damping prevents the upstream waves from destabilizing the spatial march, but results in a minimum step size below which the march becomes unstable (preventing convergence of the method) and also has the unintended consequence of damping and distorting, to differing degrees, all of the downstream waves. This has negative consequences for non-modal instabilities that are associated with an interacting group of stable modes and for flows with more rapid streamwise evolution [36]. In contrast, the numerics in OWNS are convergent and not restricted to a dominant wavelength. With no imposed dominant wavenumber, OWNS is capable of tracking the evolution of any superposition of waves and can provide solutions for arbitrary inlet boundary conditions or inhomogeneous forcing terms.

OWNS was initially developed and applied to free-shear layers [37, 38], and thereafter extended to wall-bounded flows and validated by comparison with PSE, LST, and DNS for a variety of two- and three-dimensional flows [18, 39]. The governing equations for OWNS begin with the Laplace-transformed, cross-stream-discretized, linearized, forced Navier-Stokes equations:

$$-i\omega G q' + A_{\xi, ivs}\frac{dq'}{d\xi} + Bq' = C\frac{dq'}{d\xi} + B_{\xi\xi}\frac{d^2 q'}{d\xi^2} + \mathcal{B}_p f, \tag{10}$$

where $f$ is a general primitive forcing term that is dimensionally consistent with $G\frac{\partial q'}{\partial t}$ and $\mathcal{B}_p$ is an operator that maps unit inputs into the system's state space for input-output analysis [40]. The aforementioned baseflow coefficient matrices are functions of $(\xi, \eta, \zeta)$ and have been normalized with the original coefficient matrix of $\frac{\partial q'}{\partial t}$ and further scaled by the local speed of sound $a$ such that $G = \frac{1}{a}I$. The right-hand-side terms $C\frac{dq'}{d\xi}$ and $B_{\xi\xi}\frac{d^2q'}{d\xi^2}$ correspond to the streamwise viscous quantities that can be discretized explicitly and treated as a forcing via an implicit-explicit integration scheme, such as IMEX-BDF, as described in [18]. However, since we are in the hypersonic regime, these terms will be neglected, which is equivalent to invoking the thin shear-layer approximation. Thus, we can simplify Eq. 10 and set $\mathcal{B}_p \equiv I$ (since we have no input restrictions) yielding

$$A_{\xi,ivs}\frac{dq'}{d\xi} + Mq' = f, \quad M = -i\omega G + B. \tag{11}$$

We now transform the semi-discretized primitive equation to characteristic space via the transformation

$$\phi(\xi, \eta, \zeta, t) = T_\phi(\xi, \eta, \zeta)q'(\xi, \eta, \zeta, t), \quad \widetilde{A}_{\xi,ivs} = T_\phi A_{\xi,ivs} T_\phi^{-1}, \tag{12}$$

where the rows of $T_\phi$ are the left eigenvectors of $A_{\xi,ivs}$. The discretized characteristic equation reads

$$\frac{d\phi}{d\xi} - \widetilde{M}\phi = \widetilde{A}_{\xi,ivs}^{-1} T_\phi f, \quad \phi(\xi = \xi_0, \eta, \zeta) = \phi_0, \tag{13a}$$

where

$$\widetilde{M} = -\widetilde{A}_{\xi,ivs}^{-1}\left(T_\phi M T_\phi^{-1} + \widetilde{A}_{\xi,ivs} T_\phi \frac{dT_\phi^{-1}}{d\xi}\right), \quad \phi = \begin{bmatrix}\phi_{+-}\\ \phi_0\end{bmatrix}. \tag{13b}$$

Note that subscript $0$ corresponds to the inlet plane, whereas subscripts $0, +, -$ denote the zero, plus, and minus characteristics, respectively. Eq. 13 is still exact, but cannot be solved as an initial-value problem in $\xi$ because $\widetilde{M}$ has eigenvalues of both signs. In PSE, this equation is *regularized* to damp the upstream modes, whereas in OWNS, the equation is *parabolized* by filtering out the modes with upstream group velocity. Details of transforming Eq. 13 into an approximate, well-posed, one-way equation, i.e. the OWNS differential-algebraic system of equations (DAE), can be found in [38] and [41].

## C. Arbitrary Flows

Global linear methods are the most general and are not restricted by underlying assumptions about the mean flow or the shape and path of disturbances. This means that in principle they can be used to give insight into disturbance growth and boundary layer receptivity for arbitrary three-dimensional flow fields. There are important practical caveats to employing such methods, however, including their higher computational cost due in part to the fine grid resolution required to resolve disturbances in all directions. Another practical caveat is that a user cannot simply sample all possible combinations of arbitrary disturbances to analyze the stability of a given flow field. Instead, some assumptions must be made about the type and amplitude of the input disturbance that is utilized, which is often a nontrivial choice, especially when comparing simulations with experiments. In this paper, we consider two different global methods: Adaptive Mesh Refinement Wave Packet Tracking (AMR-WPT), and forced Direct Numerical Simulation (DNS).

*1. Adaptive Mesh Refinement Wave Packet Tracking (AMR-WPT)*

The Adaptive Mesh Refinement Wave Packet Tracking (AMR-WPT) methodology is another global approach that was developed as an efficient high-fidelity transition prediction technique applicable to complex geometries [35, 42]. It has been successfully used to simulate the evolution of wavepackets for a wide range of high-speed flow conditions [43–46], and it has been demonstrated to provide a fidelity similar to conventional Direct Numerical Simulations (DNS). In this approach, a wave-packet is introduced to the laminar base flow through a pulse disturbance generated by a volume forcing term or a blowing-suction slot at the wall, as illustrated schematically in figure 2. The initial disturbance flow field is highly dependent on the spatio-temporal characteristics of the pulse disturbance and it must be chosen carefully in a way that triggers the relevant instability mechanisms. Once the disturbances are introduced, they are evolved in time using an appropriate numerical scheme and the flow field is sampled at several downstream locations to later compute growth rates and amplitudes of the disturbance evolution. One of the key features of AMR-WPT is that it employs an overset mesh technique, where the disturbance flow field is simulated separately from the baseflow utilizing an immersed boundary method [47–49]. This approach, coupled with adaptive mesh refinement, significantly alleviates some of the challenges of manual mesh generation and for tracking of small-amplitude disturbances in complex flow fields.

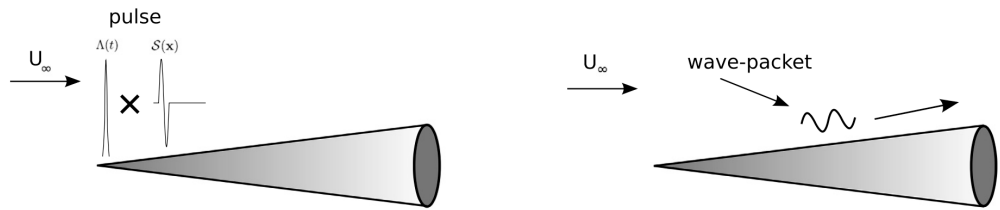

**Fig. 2 Schematic from ref. [21] showing a pulse introduced into the flow field and the subsequent convection of a wavepacket in the streamwise direction.**

In order to allow for the decomposition of the flow field into a baseflow and a disturbance flow solution, the so-called Nonlinear Disturbance Equations (NLDE) are derived from the full set of compressible Navier-Stokes equations (c.f. equation set 1). The state vector **q** is decomposed into a steady-state contribution $\bar{\mathbf{q}}$ (the baseflow solution) and an unsteady contribution $\mathbf{q}'$ (the disturbance flow solution) that retains both the linear and nonlinear terms as follows:

$$\mathbf{q} = \bar{\mathbf{q}} + \mathbf{q}' = \underbrace{\begin{bmatrix} \bar{\rho} \\ \bar{\rho}\bar{u} \\ \bar{\rho}\bar{v} \\ \bar{\rho}\bar{w} \\ \bar{\rho}\bar{E}_t \end{bmatrix}}_{\text{baseflow}} + \underbrace{\begin{bmatrix} \rho' \\ \rho'\bar{u} + \bar{\rho}u' \\ \rho'\bar{v} + \bar{\rho}v' \\ \rho'\bar{w} + \bar{\rho}w' \\ \bar{\rho}E'_t + \rho'\bar{E}_t \end{bmatrix}}_{\text{linear disturbance}} + \underbrace{\begin{bmatrix} 0 \\ \rho'u' \\ \rho'v' \\ \rho'w' \\ \rho'E'_t \end{bmatrix}}_{\text{nonlinear disturbance}}. \quad (14)$$

Similarly, the total (convective and viscous) fluxes are decomposed into steady and unsteady (linear and nonlinear) contirbutions and have the form:

$$\mathbf{E} = \bar{\mathbf{E}} + \mathbf{E}', \quad \mathbf{F} = \bar{\mathbf{F}} + \mathbf{F}', \quad \text{and} \quad \mathbf{G} = \bar{\mathbf{G}} + \mathbf{G}', \quad (15)$$

where $\mathbf{E}'$, $\mathbf{F}'$ and $\mathbf{G}'$ represent the total flux in the $x$, $y$, and $z-$direction, respectively, in Cartesian coordinates. After eliminating the pure baseflow contributions in the governing equations, the final form of the nonlinear disturbance equations can be expressed as:

$$\frac{\partial \mathbf{q}'}{\partial t} + \frac{\partial \mathbf{E}'}{\partial x} + \frac{\partial \mathbf{F}'}{\partial y} + \frac{\partial \mathbf{G}'}{\partial z} = 0. \quad (16)$$

Note that the NLDE (c.f. equation 16) are formulated in a way such that when purely linear stability investigations need to be carried out it is possible to discard the nonlinear terms, thus reducing the set of equations to the linear disturbance equations (c.f. equation 3). Solving only the linearized equations is justified for relatively low amplitude disturbances, when the nonlinear disturbance terms are small enough to be neglected (i.e., $\mathbf{q}' \ll \bar{\mathbf{q}}$). For larger initial disturbance amplitudes (i.e., $\mathbf{q}' \sim \bar{\mathbf{q}}$) and/or when the disturbances reach the nonlinear transition regime, the nonlinear disturbance terms in the governing equations become significant and need to be evaluated to accurately capture the mechanisms driving transition. A more detailed discussion of the derivation and some of the intricacies of the disturbance flow equations and its numerical solution procedure are given in [35, 50, 51].

*2. Forced Direct Numerical Simulation (forced DNS)*

Lastly, rather than solving either the linearized disturbance equations (c.f. equation 3) or nonlinear disturbance equations (c.f. equation 16), one can directly solve the full nonlinear Navier-Stokes equations (c.f. equation set 1) subject to small-amplitude unsteady forcing, a method referred to as "forced DNS." This approach has been successfully applied to analyze the stability characteristics of several different complex hypersonic flow configurations including the finned-cone [52], canonical geometries [53, 54], and in support of recent flight test experiments [55, 56].

The forced DNS approach starts with computing a steady laminar baseflow, just as in all the other linear methods. Next, small amplitude unsteady disturbances are added to the same computational grid and the perturbed baseflow is evolved by solving the nonlinear equations in time using an appropriate time-accurate computational method. As with the other global methods, the introduced disturbances are free to evolve along any path in three dimensions and can excite boundary-layer instabilities with sufficiently small amplitudes to remain in the linear regime.

Aside from choosing an appropriately small initial disturbance amplitude, the main challenge with this approach is in utilizing an appropriate grid and numerical scheme to not artificially attenuate or amplify the disturbances that are introduced. In particular, since forced DNS considers the evolution of the perturbed baseflow rather than only the perturbations, the use of low-dissipation numerics are critically important to obtaining accurate solutions, as is discussed in section III.D.5. Furthermore, the approach allows the flexibility to simulate beyond the linear regime and implicitly includes any small modifications to the baseflow by the perturbations, which may become important if secondary instabilities are present.

Once the forced DNS simulations are run long enough for the disturbances to at least pass through the outflow of the computational domain, information about the stability characteristics of the flow field can be extracted. This is typically done in terms of an amplitude or N-factor by carefully examining the spatial variation of any of the primitive variables, e.g., the growth of pressure fluctuations along the wall.

## III. Computational Methods

### A. Computational Grids

*1. Baseflow Grids*

Gridding software LINK3D was used to generate high-quality structured grids for the finned-cone laminar mean flow. The CAD model of the geometry was provided through collaboration with Purdue University. Four grids were initially constructed with increasing azimuthal ($\phi$) resolution to check for convergence of the mean flow. Relevant parameters for each grid are given in table 3. Baseflow grids I and II were quarter symmetry grids with 102 M and 111 M cells, respectively, and average azimuthal spacing at the end of the domain of $\triangle \phi = 0.30°$ and $\triangle \phi = 0.21°$.

Table 3  Baseflow grid parameters.

| Baseflow grid | $x$-domain (m) | $\phi$-domain | Total Cells | Wall-normal points | $\triangle \phi$ at $x = 0.4$ m |
|---|---|---|---|---|---|
| I | $0 < x < 0.41$ | $0° < \phi < 90°$ | 102 M | 270 | 0.30° |
| II | $0 < x < 0.41$ | $0° < \phi < 90°$ | 111 M | 270 | 0.21° |
| III | $0.08 < x < 0.41$ | $0° < \phi < 60°$ | 146 M | 270 | 0.13° |
| IV | $0.08 < x < 0.41$ | $0° < \phi < 60°$ | 388 M | 324 | 0.07° |

To minimize the computational cost of obtaining highly-resolved laminar flow fields, an "interpolated subdomain" technique was employed. First, a precursor simulation is performed using a relatively coarse grid that covers the entire geometry; this grid adequately resolves the flow over the nose and conical region ahead of the fin, but inadequately resolves the fin interaction region. Next a highly resolved subdomain grid is generated beginning slightly upstream of the fin root; this subdomain contains only the fin interaction region. Finally the solution (including gradient information) from the precursor simulation is interpolated onto the boundaries of the subdomain grid and the solution in this region is computed. This approach is similar to the method used in unsteady DNS with precursor computations (see, e.g., [56, 57]) and allows for increased spatial resolution of the laminar mean flow utilizing fewer grid points. Following this approach, baseflow grids III and IV were subdomain grids with 146 M and 388 M cells, respectively, within a domain that extended from 0.08 m$< x <$0.41 m and $0° < \phi < 60°$. For all mean flow grids, the spacing at the first wall-normal cell in wall units corresponded to $\Delta y^+ \ll 1$ everywhere in the domain of interest.

*2. Stability Analysis Grids*

Each linear stability code has its own unique way of utilizing a given baseflow, which typically includes interpolating the mean flow onto a new computational grid for the stability analysis. For the current work, an attempt was made to standardize this to some degree by utilizing a portion of the finest resolution baseflow (grid IV in table 3) that excluded the fin; this choice for analysis was based on experimental observations of transition onset occurring along the acreage of the cone [25]. A domain was selected spanning from $15° < \phi < 60°$, 0.09 m$< x <$ 0.40 m, and $\eta < 0.01$ m, where $\phi$ is measured from the fin and $\eta$ is the wall-normal coordinate; this is referred to as the "common mean flow domain" and is shown schematically in figure 3 along with grid parameters in table 4.

Table 4   Common Mean Flow Grid Parameters.

| $x$-domain (m) | $\phi$-domain | $\eta$-domain (m) | Total Cells | Wall-normal points | $\Delta\phi$ at $x = 0.4$ m |
|---|---|---|---|---|---|
| $0.09 < x < 0.40$ | $15° < \phi < 60°$ | $\eta < 0.01$ | 10 M | 250 | $0.08°$ |

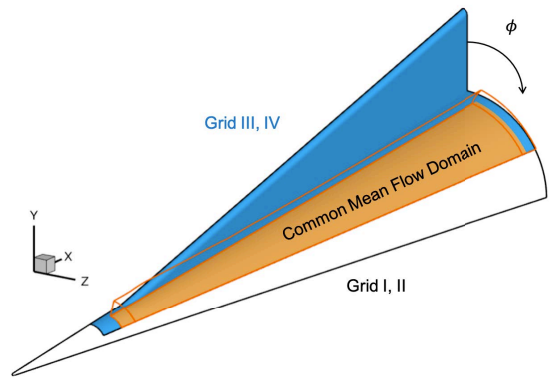

Fig. 3   Schematic of computational grid domains corresponding to tables 3 and 4

In figure 3, the blue region represents the interpolated subdomain grid extents given in table 3 (baseflow grids III and IV) and the orange region represents the common mean flow domain, which included $100 \times 400 \times 250$ (axial, spanwise, wall-normal) grid points and was clustered in the spanwise direction to properly capture spanwise gradients within the vortical structures. The majority of this common domain also included grid cells underneath the shock except for a small region near the fin-cone junction that crosses the shock.

**B. Baseflow Solver**

Computations of the common laminar mean flow were executed with US3D [58], an unstructured finite-volume Navier-Stokes compressible flow solver. First-order implicit time integration with the data-parallel line relaxation method is used to march the solution to steady state. Inviscid fluxes are computed with a second-order accurate kinetic energy consistent scheme, which switches between high dissipation, stable fluxes near shocks and low dissipation, high accuracy fluxes elsewhere in the domain using a switch based on the Mach number [59].

**C. Flow Conditions**

Inflow conditions for the baseflow computations correspond to experimental conditions from experiments at the Purdue University Boeing Air Force Mach 6 Quiet Tunnel (BAM6QT) [25], which is summarized in table 5. An isothermal wall is imposed and a calorically and thermally perfect gas (air) is used for all simulations.

Table 5   Inflow conditions corresponding to BAM6QT experiment

| Reynolds No. (m$^{-1}$) | Velocity (m/s) | Static temp. $T_0$ (K) | Static Pressure (Pa) | Wall Temp. $T_w$ (K) |
|---|---|---|---|---|
| 8.4×10$^6$ | 865.52 | 51.78 | 482.10 | 309.5 |

**D. Stability Computations**

The remainder of this section gives details of the computational setup for each of the codes used to analyze the stability of the common mean flow for the finned-cone.

*1. LASTRAC LST implementation*

The NASA Langley Stability and Transition Analysis Code (LASTRAC) was developed to provide a freely available tool for routine boundary-layer transition prediction on three-dimensional geometries [60]. The code has been developed over nearly two decades and has a range of capabilities that implement both LST and (linear and nonlinear) PSE analysis. It has also been validated against a number of different flow configurations and used to investigate boundary-layer instabilities in several different hypersonic flight test experiments [61–63].

In this work, LASTRAC is used for local LST computations for the finned cone. The common mean flow was converted into the required binary format and a series of 200 disturbance paths with 500 points along each were generated for the analysis. These disturbance paths were chosen such that an even spanwise distribution of paths would be achieved at a distance 70% of the axial length of the subdomain being analyzed. Furthermore, the analysis was restricted to non-oblique instabilities ($\beta = 0$) with a phase speed between 40–100% of the edge velocity and frequencies in the range of 5-500 kHz.

Standard practice in LST analysis for hypersonic flows is to use the boundary-layer edge quantities to define a marching path for the analysis of disturbances. Typically, a marching path utilizes the velocities at a boundary-layer edge defined at 99.5% of the freestream total enthalpy. An initial set of LASTRAC runs was conducted using enthalpy-based disturbance paths for the analysis, but it was found that these paths were slightly misaligned with vortex structures later shown by other methods to be important to the transition process. After some experimentation, it was found that using a disturbance path based on the freestream velocity magnitude resulted in paths that more closely followed the vortex structures. Therefore, the results shown here are for disturbance paths utilizing an edge definition of 99.5% of the freestream velocity magnitude (under the shock). Further details on the numerics and other user settings within LASTRAC are given in ref. [60].

*2. LSTPACK BiGlobal and planar PSE implementation*

The general BiGlobal and planar PSE approaches described in section II.B.1 are implemented in a code called LSTPACK originally developed at Sandia National Laboratories. Discretization of the planar PSE equations with respect to the spanwise ($\zeta$) and wall-normal ($\eta$) directions is accomplished using fourth-order central finite differences in the interior of the domain and second-order biased finite difference stencils at the boundaries. Streamwise discretization for the planar PSE marching employs a first-order accurate, implicit scheme. The PSE march is subject to the constraint:

$$\left\langle \frac{\partial \hat{\mathbf{q}}}{\partial \xi}, \hat{\mathbf{q}} \right\rangle = 0, \tag{17}$$

which minimizes the variation of the shape function $\hat{\mathbf{q}}$ along the streamwise $\xi$ direction. The angled brackets used above denote the inner product, which is defined for two arbitrary vector functions $\mathbf{f}$ and $\mathbf{g}$ by:

$$\langle \mathbf{f}, \mathbf{g} \rangle \equiv \int_\eta \int_\zeta \mathbf{g}^H \mathbf{f} d\eta d\zeta \tag{18}$$

where superscript $H$ denotes the conjugate transpose. The constraint imposed by Equation 17 uniquely determines the decomposition of the disturbance $\mathbf{q}'$ into its wave part $\alpha$ and its shape function $\hat{\mathbf{q}}$.

The BiGlobal stability analysis is performed by solving the eigenvalue problem of equation 9. Following discretization, the dimension of the matrices $\mathbf{L}$ and $\mathbf{R}$ is $N = 5 \times N_\eta \times N_\zeta$, where $N_\eta$ and $N_\zeta$ are the number of grid points in the wall-normal and spanwise directions. The size of these matrices can become very large. Consequently, conventional dense eigenvalue algorithms, such as the QR algorithm, are unacceptably expensive. Instead, the implicitly-restarted Arnoldi method as implemented in the ARPACK software package [64] is employed to seek only the most unstable eigenvalues of Equation 9 rather than the full spectrum. This procedure brings the eigenvalue search to an acceptable level of computational expense.

Both the BiGlobal eigenvalue problem (equation 9) and the planar PSE marching (equation 8) require the solution of large linear algebra problems. These linear algebra problems are formulated using sparse matrices to minimize storage and are solved in parallel using the MUMPS (MUltifrontal Massively Parallel Solver) software library [65]. It is well-known that the PSE are unstable for small marching steps unless pressure gradient suppression is used to remove residual ellipticity from the equations [66]; however, in the current analysis the marching steps were large enough that no evidence of such instability was observed.

The wall boundary condition for the stability analysis was a no-slip wall with zero temperature fluctuations. At the spanwise and top edges of the stability analysis domain, a Dirichlet boundary condition was enforced through the use of a sponge layer that caused instabilities to damp out near the simulation boundary. Except very near the fin-root, the vortex instabilities analyzed in this paper were confined to the interior of the analysis domain and thus were minimally influenced by these boundary sponge layers.

A typical stability analysis utilizing LSTPACK proceeds according to the following steps:

1) An initial guess of the BiGlobal eigenvalue is provided (as described in Section II.B.1)
2) The Arnoldi method is used to find the most unstable eigenvalues of Equation 9 that lie closest to the guess.
3) The wavenumber and eigenfunction from the BiGlobal analysis are provided as an initial condition for the planar PSE march.
4) At each planar PSE marching step, the solution at the downstream location is determined by solving Equation 8 with an implicit streamwise discretization. This is done iteratively while applying Newton's method to drive Equation 17 to zero.
5) At each planar PSE marching step, iterations are performed until the convergence criterion $|\Delta\alpha_c/\alpha_c| < 10^{-5}$ is reached, where $\Delta\alpha_c$ is the change in wavenumber $\alpha_c$ between iterations.

For the results included in this paper, planar PSE marches were initiated with BiGlobal solutions at five different axial locations between $x = 0.09 - 0.24$ m and spanned a frequency range of 1-400 kHz. The BiGlobal stability grid utilized had 250 points in the wall-normal direction and 400 points in the spanwise direction clustered about $\phi = 32°$ near the laminar vortex. Increasing the grid density beyond this did not significantly alter the results.

*3. CSTAT planar PSE and OWNS implementation*

The Caltech Stability and Transition Analysis Toolkit (CSTAT) is a comprehensive stability code capable of performing LST, spatial BiGlobal (SBG), OWNS, PSE, and global analyses for boundary and free-shear layers in generalized, non-orthogonal, body-fitted curvilinear coordinates and user-defined fluid properties. The code can solve for one or two inhomogeneous cross-stream directions and has been validated by comparison with PSE, LST, and DNS for a variety of two- and three-dimensional flows [1, 18, 39]. Recently, the OWNS algorithm has been extended to include an input-output optimization framework that is capable of efficiently calculating (linear) worst-case disturbances that lead to the fastest transition to turbulence [41, 67–69].

Due to the increased computational cost of OWNS over PSE, the common base flow was truncated in the spanwise ($\zeta$) and wall-normal ($\eta$) directions to allow the same cross-stream resolution in both PSE and OWNS calculations. This domain truncation was found to have minimal impact on the overall growth rate and thus we use the truncated domain for all CSTAT calculations presented in this paper. Even with this added efficiency, the OWNS calculations still required substantial time and memory (RAM) to invert the DAE matrix at each streamwise station via the lower-upper (LU) decomposition. To circumvent this, we developed a hybrid computational approach in which the LU decomposition was performed on the DAE matrix constructed with second-order cross-stream discretization with reduced-order boundary closure, which significantly increased the sparsity of the matrix and thereby reduced the memory cost. This operation was performed using the Intel® oneAPI Math Kernel Library PARDISO package [70] to parallelize the LU decomposition. The inverted system and corresponding solution then served as a preconditioner matrix and guess, respectively, for solving the full linear system using the generalized minimum residual (iterative) method from MATLAB. This hybrid approach reduced the total computational time by $\approx 50\%$ at each streamwise station.

The computational setup of the PSE and OWNS calculations are presented in Table 6, where $N_b$ is the number of recursion parameters used to filter upstream waves from the OWNS marches and $N_b = 10$ was found to give recursion-parameter-converged results. The PSE and one of two coarse-grid OWNS calculations ($N_\xi = 2001$) are initialized using the same SBG eigenfunction. The second coarse-grid OWNS computation ($N_\xi = 1232$) is initialized using randomized inlet forcing, i.e. forcing of each state variable at every grid point randomly drawn from a normal distribution. This is done to excite all wavenumbers at a given frequency and to trigger the gamut of potential instability mechanisms. No-slip and isothermal boundary conditions are imposed at the wall ($u' = v' = w' = T' = 0$), and thus we solve the linearized continuity equation for $\rho'$. At the upper boundary, we impose 1D inviscid Thompson characteristic boundary conditions [71] and at the spanwise boundaries we enforce symmetry conditions.

**Table 6  Computational setup of PSE and OWNS marches at $f$ = 250 kHz.**

| Solver | $N_\xi$ | $N_\eta$ | $N_\zeta$ | $N_b$ | $x_{min}$ [m] | $x_{max}$ [m] | Inlet BC |
|---|---|---|---|---|---|---|---|
| PSE | 627 | 225 | 301 | - | 0.090 | 0.400 | SBG |
| OWNS | 2001 | 225 | 301 | 10 | 0.090 | 0.400 | SBG |
| OWNS | 1232 | 225 | 301 | 10 | 0.090 | 0.280 | Random |
| OWNS | 6001 | 225 | 301 | 10 | 0.245 | 0.333 | OWNS ($N_\xi = 2001$) |

The cross-stream directions are discretized with fourth-order central finite-difference schemes with summation-by-parts [72] boundary closure. Both the streamwise and spanwise coordinates are uniformly distributed, with grid-stretching in the wall-normal direction [73]. The OWNS marches are integrated in the streamwise direction via second-order backward differentiation formula. For the purpose of comparing OWNS predictions to the other linear methods in this paper, the highest-resolution OWNS computation ($N_\xi = 6001$) was initialized from a coarser simulation ($N_\xi = 2001$) starting at $x = 0.245$ m. This was done after earlier results [1] that included a coarser OWNS solution were found to be not fully streamwise-grid converged. Further investigation revealed that $N_\xi = 6001$ was sufficient streamwise spacing to agree closely with the planar PSE predictions, which were known to be stability-grid converged. Performing the OWNS calculation at this highest streamwise resolution for the entire domain would be computationally prohibitive, however, due to the limits of the current implementation of the method.

*4. CHAMPS AMR-WPT implementation*

The AMR-WPT methodology described in section II.C.1 is implemented as part of the Cartesian High-order Adaptive Multi-Physics Solver (CHAMPS). To solve the disturbance flow equations (c.f. equation 16) a modified fifth-order accurate finite-difference WENO [74] scheme was used for the convective terms and second-order centered finite-difference operators were used for the viscous terms. A fourth-order accurate explicit Runge-Kutta time-integration scheme is used to advance the solution in time.

As mentioned previously, the AMR-WPT method employs a dual mesh overset approach, consisting of a baseflow mesh and a disturbance flow mesh, where the baseflow solution is interpolated onto the disturbance flow mesh at every re-meshing step. The overset interpolation approach has been optimized to a point where its impact on the computational performance is negligible as refinement is typically only applied every 50-100 timesteps. The timestep interval when refinement occurs is typically determined based on the fast acoustic wave-speed and the block size used in the simulations. The disturbance flow mesh is usually a block-structured Cartesian grid organized in an octree data structure for efficiency but also supporting multi-block generalized curvilinear meshes.

Tracking of disturbances is performed on the basis of a mesh refinement sensor function and the AMR algorithm is designed to determine which regions of the flow require high/low grid resolution and refines/derefines the mesh accordingly. Various mesh refinement sensor functions based on different disturbance flow quantities can be used to control the mesh refinement and derefinement. Numerical tests performed in prior work [45, 50] demonstrated that the best tracking was obtained by basing the mesh refinement sensor function on pressure, velocity and vorticity disturbances, as well as the gradient of pressure fluctuations. Based on this prior experience, the same composite sensor function has been used in the present work.

In order to simulate the disturbance flow field around complex geometries on a Cartesian mesh, a previously developed higher-order immersed boundary method (IBM) [47–49, 75, 76] has been modified to be applicable for the solution of the nonlinear disturbance equations. Fig. 4 displays the finned cone treated as an immersed boundary within the block-structure Cartesian computational domain and an instantaneous wireframe view of the adaptive grid, with the finest levels tracking the disturbance flow field. In the current work, the common mean flow (as defined in section III.A.2) was extrapolated along its boundary faces to fit within the baseflow Cartesian domain, effectively serving as sponge layers outside the interior region of interest to the stability analysis. More details about the IBM approach to solving the nonlinear disturbance equations as well as extension of the IBM-AMR method to deal with mesh refinement along a sharp immersed boundary are provided in refs. [?] and [77], respectively.

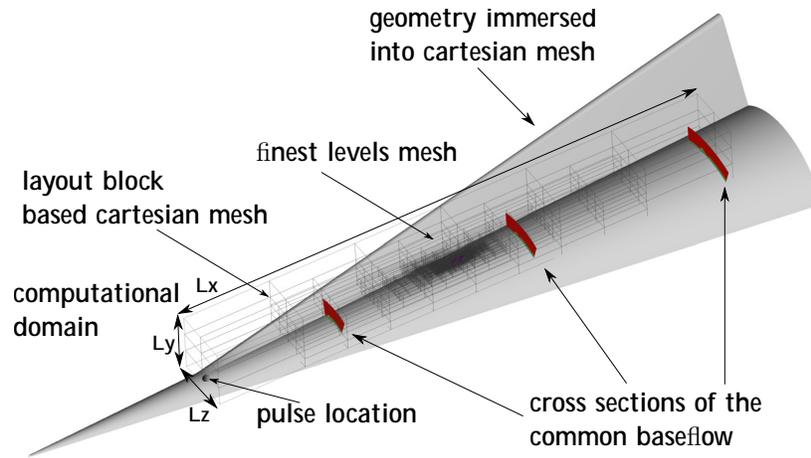

**Fig. 4    Finned Cone treated as an Immersed Boundary in a Cartesian Box (depicted as a grey wireframe). The black wireframe displays an instantaneous grid generated by the adaptive algorithm around the disturbance.**

The disturbances within the computational domain can be introduced in several ways and can excite different types of disturbances, e.g., acoustic, vorticity and entropy waves. For the present AMR-WPT simulations on the finned cone, the disturbance field is introduced by means of a volume forcing short duration pulse, where forcing is applied to the y-momentum equation and then subsequently added to the right-hand-side, similar to the method adopted in ref. [78]. The pulse has a Gaussian shape in the wall-normal direction and sinusoidal distribution in streamwise direction. During the short pulse duration, the pulse amplitude increases over time according to a Gaussian distribution function. For the AMR-WPT simulations used to compare with the other linear methods, the pulse period was set equal to $4 \times 10^{-6}$ s.

*5. US3D forced DNS implementation*

Forced DNS simulations were conducted using the same code (US3D) as was used to generate the common mean flow. However, due to more strict requirements on gridding and numerics for simulating added disturbances, a new baseflow was required using a slightly different approach than previously described in section II.A. 2. First, the finned-cone domain was truncated from the nose to $x = 0.05$ m (upstream of the fin) at an angle normal to the cone surface; this results in a conical inflow surface over just the upstream blunt cone portion of the geometry. With this truncated domain, a steady, axisymmetric solution was computed using fourth-order low-dissipation spatial fluxes [53] with dissipative shock capturing fluxes that are active at shocks to provide numerical stability [59].

With the steady, axisymmetric blunt cone solution, a wall-normal solution profile was extracted at a location that coincided with the ghost cells of the downstream finned-cone subdomain; the resolution of this subdomain grid is given in table 7 for both a 'coarse' and 'fine' grid utilized in this paper. Based on prior work [27, 52] a small fillet was also added to the fin-cone junction to facilitate the gridding in the corner region for both grids. The axisymmetric blunt cone solution profile was then interpolated onto the finned-cone subdomain inflow ghost cells and a new baseflow solution was time-marched to steady-state using the same low-dissipation flux and shock capturing scheme described above. Once the laminar solution reached steady state, small amplitude unsteady disturbances were introduced at the inflow of the finned-cone subdomain to excite boundary-layer instabilities and the solution was advanced in time using a second-order time-accurate implicit method [52]. This approach is similar to the work of ref. [56] and is illustrated schematically in figure 5.

Table 7   Forced DNS finned cone subdomain grid resolution

|  | Cells | Wall Faces | $\Delta x$ (mm) | $\Delta \phi$ (deg) |
|---|---|---|---|---|
|  |  |  | $\phi = 35$ degrees |  |
| Coarse | 158M | 0.54M | 0.276 | 0.19 to 0.29 |
| Fine | 347M | 1.14M | 0.197 | 0.09 to 0.12 |

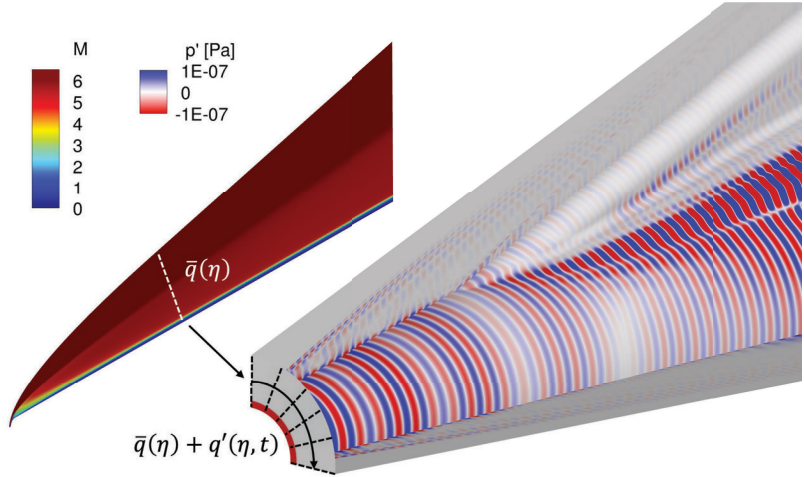

Fig. 5   Illustration of the baseflow generation and forced DNS approach for the finned cone.

The inflow disturbances had a temporal frequency of $f = 247$ kHz and were introduced into the boundary layer as perturbations to the pressure field with the following form:

$$p'(\eta, t) = \begin{cases} \frac{1}{2}(1 + \cos(\pi \frac{\eta}{\delta})) \cos(2\pi f t) & \text{if } \eta \leq \delta \\ 0 & \text{if } \eta > \delta \end{cases} \qquad (19)$$

with $u' = v' = w' = T' = 0$ and $\rho'/\bar{\rho} = p'/\bar{p}$. To ensure that the disturbances remained in the linear regime ($\mathbf{q}' \ll \bar{\mathbf{q}}$), the disturbance amplitude at the inflow was set to $|p'|/\bar{p}_w = 10^{-9}$. Time-series data was collected throughout the unsteady simulation in order to compare disturbance profiles and N factors with results from the other linear stability analysis methods.

## IV. Laminar Baseflow

Steady-state laminar computations were performed for each of the four baseflow grids outlined in table 3 of section III.A.1. Figure 6 shows a three-dimensional visualization of the flow field utilizing baseflow grid IV and figure 7a shows a comparison of the computed streamwise velocity at two different axial stations ($x = 0.15$ m and $x = 0.35$ m) for grids I-IV. It is clear from figure 7a that the mean flow is highly sensitive to the grid resolution, particularly in the spanwise distribution of points used to capture the rollup of the streamwise-aligned vortex along the acreage of the cone. This becomes most apparent downstream, as is shown at $x = 0.35$ m, where the solution utilizing the lowest resolution grid (grid I) overly dissipates this vortex feature and it does not form a well-defined crest as can be seen in the highest resolution grid (grid IV). As discussed later in the paper, the mean flow in this acreage region turns out to be highly sensitive to the numerics of the computation, namely the grid resolution and spatial flux accuracy. Significant effort was made to resolve the main laminar flow features in this acreage region of interest. Ultimately, the solution utilizing grid IV with second-order fluxes was selected to compare the linear stability methods; the common mean flow discussed in section III.A.2 was extracted from this grid IV solution.

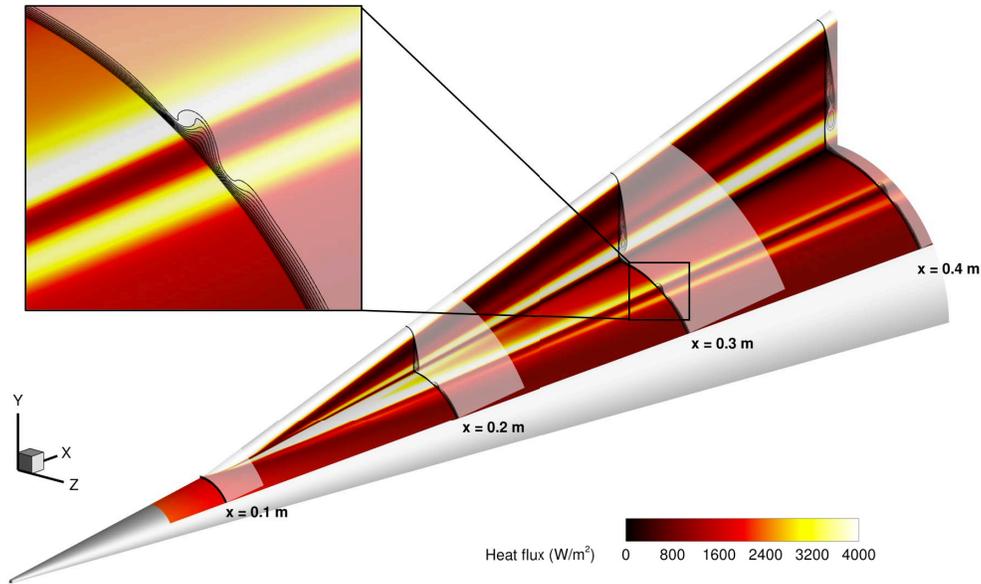

**Fig. 6 Visualization of the laminar mean flow utilizing grid IV of table 3. The inset in the image highlights part of the region of interest for the stability computations. Streamwise velocity contour slices are shown at $x = 0.1, 0.2, 0.3, 0.4$ m. Colored contours correspond to surface heat flux. Freestream conditions are given in table 5.**

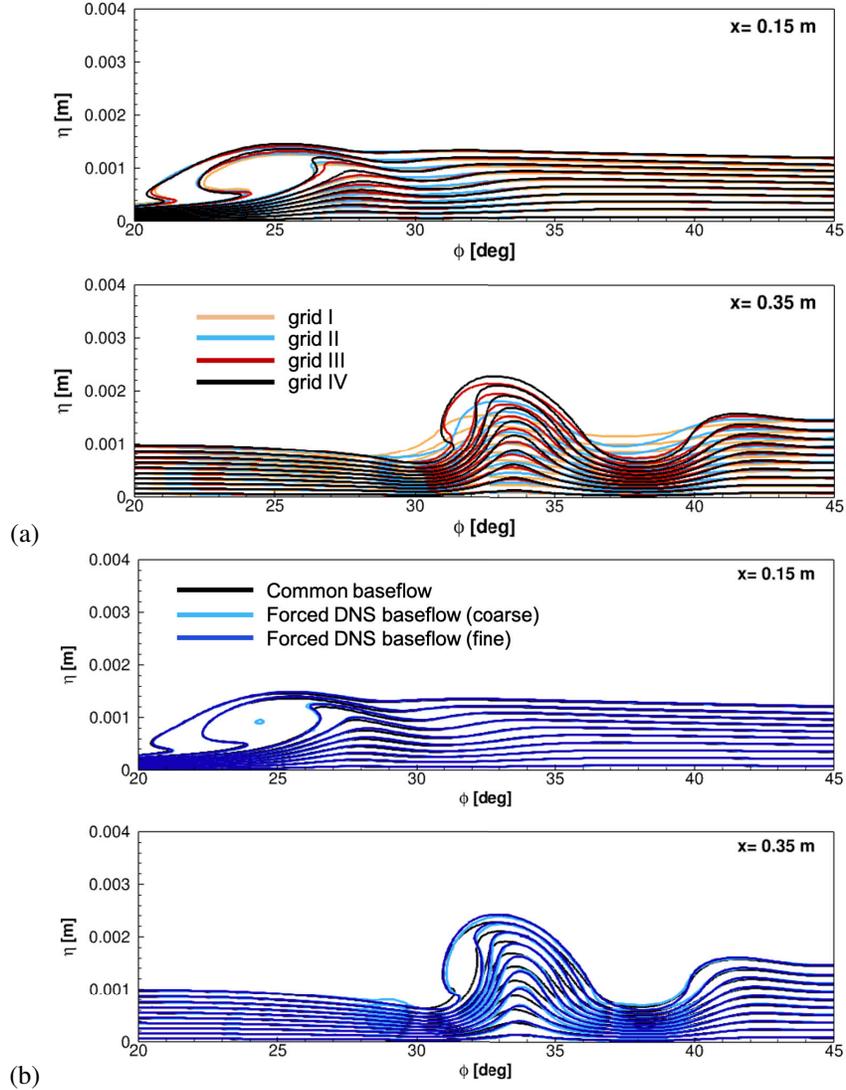

**Fig. 7  Laminar mean flow comparison of streamwise velocity ($\bar{u}$) at two axial stations, $x = 0.15$ m and $x = 0.35$ m, for (a) grids I-IV and (b) the common baseflow and the forced DNS baseflow with 'coarse' and 'fine' grids.**

As was mentioned in section III.A.6, the forced DNS simulations utilized a slightly modified grid as well as fourth-order numerics for computing the baseflow and evolving small amplitude disturbances. Figure 7b shows a comparison of the streamwise velocity contours for the common mean flow and the DNS baseflow utilizing either a 'coarse' or 'fine' grid (as defined in table 7) for the same two axial stations. Qualitatively, the baseflow solutions in figure 7b look quite similar, with only modest discrepancies appearing far downstream. These differences are discussed further within the context of the stability results that are presented in the next section. It is important to emphasize that the main focus of this paper is on comparing the linear stability methods, i.e., the relative predictions of the methods to each other and experiment for a given baseflow; this limited the need for further efforts to examine convergence of the mean flow beyond what is presented in this section.

## V. Stability Computations

This section presents results of the stability computations discussed in the prior sections as applied to the finned cone laminar baseflow described in section IV.

### A. Most Unstable Mode

For any boundary layer stability problem, two key questions that an analyst tries to answer are: (1) what disturbance characteristics (frequency, phase speed, etc.) excite boundary-layer instabilities, and (2) what is the most unstable disturbance that likely leads to transition? For the finned-cone problem at hand, we first apply LST analysis to the laminar baseflow as an initial quick look at trying to answer these questions. Figure 8a shows the LST N-factors plotted against axial distance ($x$) and angle away from the fin ($\phi$) as computed using LASTRAC for the case with disturbance paths based on velocity magnitude that was described in section III.D.1. This N-factor contour represents a compilation of the growth of non-oblique ($\beta = 0$) disturbances computed by LST along individual line marching paths. Additionally, a vertical line has been added to the figure to indicate the axial location of transition onset ($x_{tr} \approx 0.31$m) that was observed in the quiet tunnel experiments for this condition based on surface heat flux measurements [25].

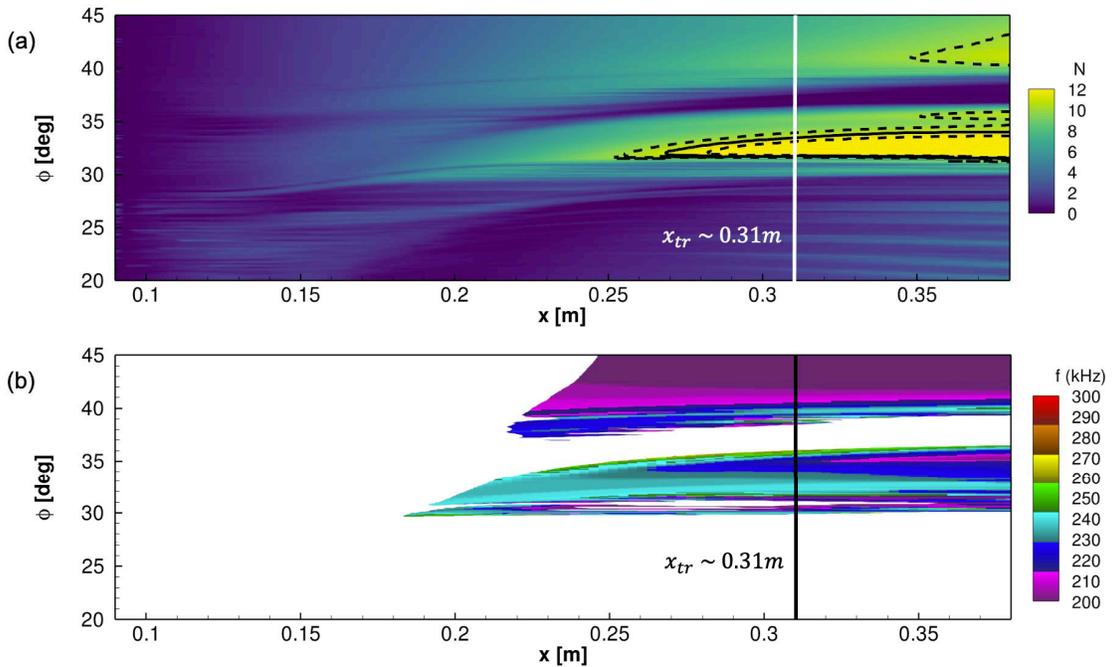

**Fig. 8** LST (a) N-factors with a solid black contour line drawn for $N = 11$ and dashed black contours lines for $N = 10$ and $N = 12$ and (b) corresponding non-oblique ($\beta = 0$) instability frequencies for $N \geq 5$ from the LASTRAC analysis.

In Figure 8a, there are two main regions with large N-factors that appear downstream. In between $\phi = 30 - 35°$, N-factors as high as 18 are found by the end of the domain. Interestingly, this region corresponds to the location of the dominant laminar vortex structure formed by the fin-cone interaction region. This region is also where there is a strong spanwise variation in the baseflow due to the vortex rollup, making the LST analysis and its underlying assumption of locally parallel flow suspect. In the outboard regions of the body, i.e., $\phi > 40°$, there is much less spanwise distortion of the baseflow (at least upstream of $x_{tr}$) and N-factors between 9–13 are found. Figure 8b shows the corresponding frequencies associated with local N-factor of $N \geq 5$. The frequencies for high N-factors between $\phi = 30 - 35°$ are approximately 220-260 kHz, while the frequencies for $\phi > 40°$ are between 140–220 kHz. At this stage in the analysis, it is reasonable to postulate that the outboard instabilities detected by LST are likely second mode, since the frequencies are in the range of what would be expected for a simple sharp cone and the mean flow appears within the limits of locally parallel flow in this region. However, the region in and around the laminar vortex requires additional work utilizing a higher fidelity stability analysis method to build confidence in understanding the results.

Aside from the LST analysis, only planar PSE and AMR-WPT analyses conducted for this paper included some variation in the temporal frequency of the initial disturbance. The planar PSE analysis did so by a manual search over a limited range of possible frequencies and phase speeds and the AMR-WPT analysis did so by nature of the initial pulse disturbance utilized, which includes some finite bandwidth of frequencies and spanwise wavenumber content. Figure 9 shows the maximum surface N-factor computed from the LSTPACK planar PSE analysis for the finned-cone.

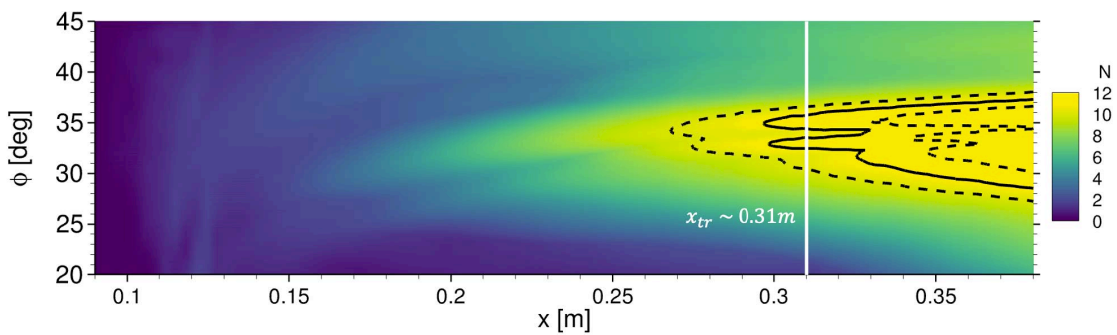

**Fig. 9 Planar PSE maximum surface N-factor as defined in equation 20 computed by LSTPACK. A solid black contour line is drawn corresponding to $N = 11$ and dashed black contours lines are at $N = 10$ and $N = 12$.**

The surface N-factors shown in figure 9 are computed using the pressure eigenfunction from the planar-PSE analysis evaluated at the wall defined as:

$$N = ln\left(\frac{|\hat{p}_w(x,\phi)|}{|\hat{p}_{w,max}(x_0)|}\right) \quad (20)$$

where $\hat{p}_w(x,\phi)$ is the amplitude of the pressure fluctuation at the wall at axial location $x$, and $\hat{p}_{w,max}(x_0)$ is the maximum amplitude of the pressure fluctuation at the wall at the initial location, $x_0$. The absolute value is used to give the magnitude of the complex shape function. This definition fixes $N = 0$ at the starting location of each PSE march and therefore does not necessarily correspond to the neutral point of any one unstable mode; however, the use of multiple PSE marches places the start of one of these marches close to the neutral point. Qualitatively comparing figure 9 with figure 8a, there are some coarse similarities in how the largest N-factor appears within the $\phi = 30 - 35°$ region and starts to grow rapidly downstream of $x \approx 0.2$ m. However, the similarities end there and it can be seen that there are large regions of instability growth that appear in the planar PSE analysis that are not seen in the LST analysis, e.g., for $\phi < 30°$. To determine the most unstable mode from the planar PSE analysis, the mode with the maximum surface N-factor at the end of the computational domain was identified. This dominant mode was in the form of a fluctuation that sits near the crest of the laminar vortex closest to the fin. Figure 10 shows a three-dimensional reconstruction of this dominant vortex mode for the finned-cone flow field analyzed. The mode oscillates at a frequency of 247 kHz and is visualized in the figure by black and white isosurfaces of streamwise velocity fluctuations ($u'_r = \mathbb{R}(u')$). Also shown for reference are contours of surface heat flux and an axial slice of the mean streamwise velocity at different axial stations from the common baseflow. The instability can be seen to follow a path aligned with the main laminar vortex that is generated by the shockwave-boundary-layer interaction initiated at the fin-cone junction upstream.

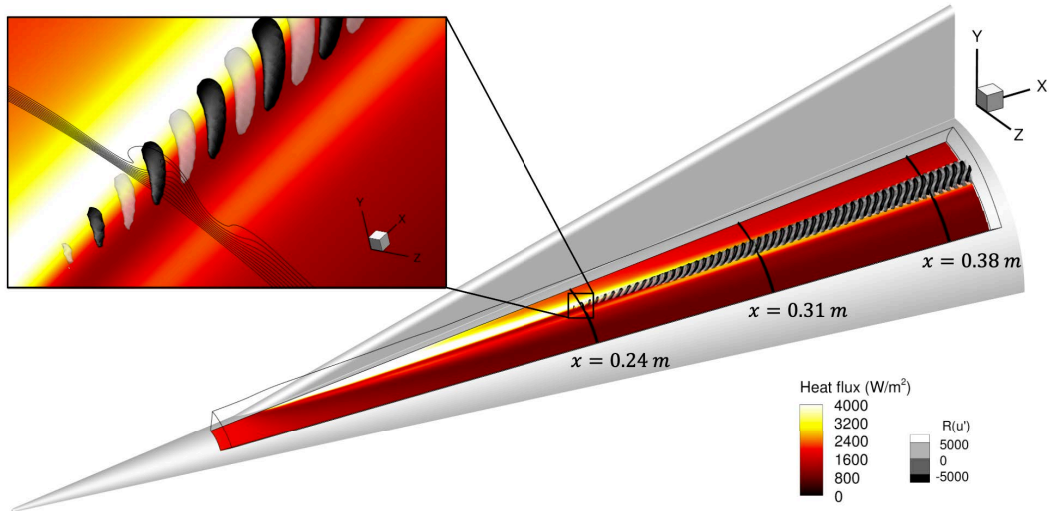

**Fig. 10 Visualization of the planar PSE prediction for the most unstable mode occuring at a frequency of approximately 250 kHz by LSTPACK. Isosurface contours of streamwise velocity fluctuations are shown in black and white for $\mathbb{R}(u') = \pm 5000$, respectively. Color contours correspond to the computed surface heat flux within the extents of the common mean flow domain.**

As a quick check of the planar PSE results, the temporal frequency content of the pulse disturbance utilized in the AMR-WPT analysis was examined. Figure 11 shows the amplitude growth (along a fixed azimuth) of different frequencies included in the pulse disturbance as it propagates through the computational domain. The band of frequencies in the range of 220 – 280 kHz are highlighted to show that they account for the largest amplitude growth within the region of interest, with the frequency at approximately 250 kHz showing the largest growth, consistent with the planar PSE analysis.

Additionally, figure 12 shows a schematic that depicts the initial pulse location as well as axial contours of the amplitude of the streamwise velocity fluctuations occurring at 250 kHz throughout the domain. It can be seen that the structure of these velocity fluctuations as well as their position in the flow field are qualitatively similar to the depictions of the unstable vortex mode shown in figure 10. Further comparisons of the vortex mode shape will be shown later along with computations utilizing the other linear methods.

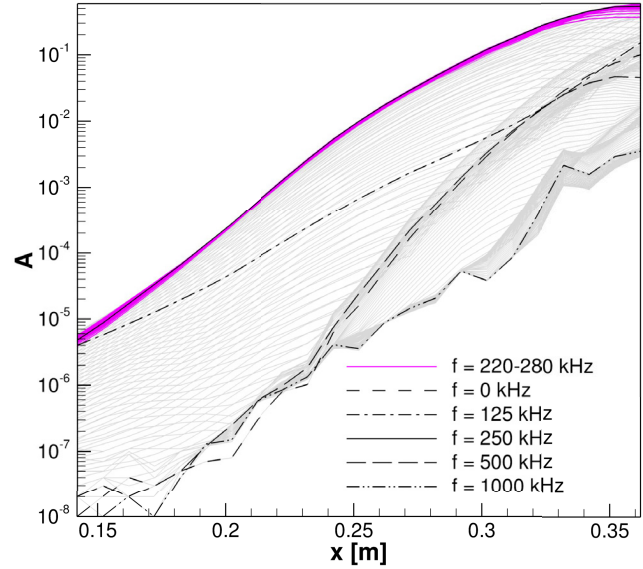

**Fig. 11   Amplitude growth along a fixed azimuth of the different frequency components included in the disturbance pulse used in AMR-WPT analysis.**

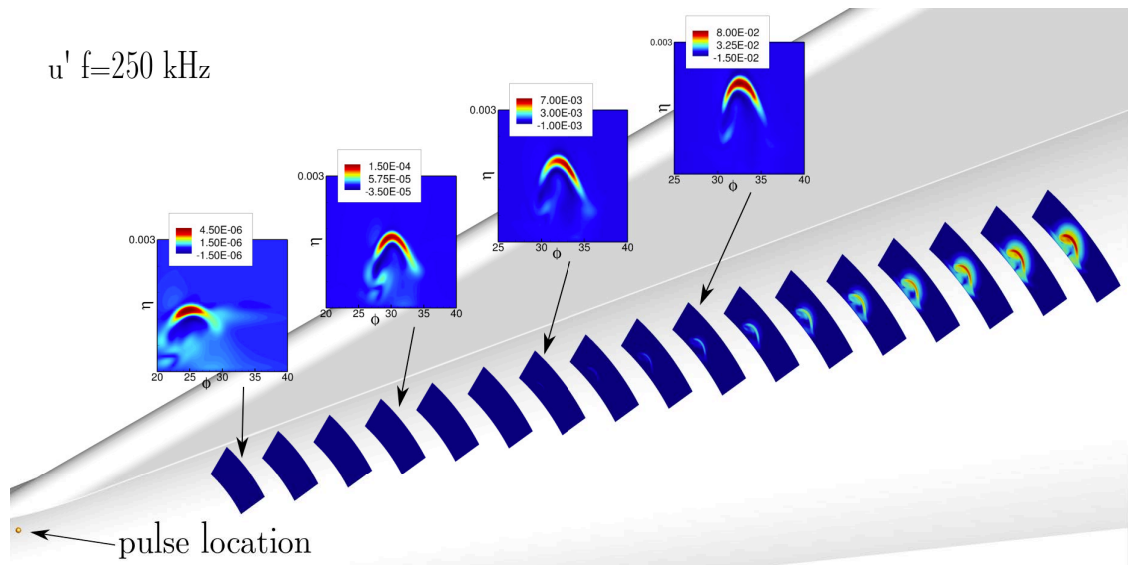

**Fig. 12   Schematic showing the initial location and downstream evolution of the pulse disturbance from the AMR-WPT analysis. Color contours correspond to the amplitude of the streamwise velocity fluctuations ($u'$) for only the 250 kHz component of the disturbance amplitude as it is tracked through the domain.**

## B. N-factor Comparison

As mentioned in the introduction, N-factor predictions such as the ones shown in figures 8 and 9 are standard practice in linear stability analysis. However, it is well known that such predictions have major shortfalls. One of the challenges is in simply defining the N-factor for any of the methods outside of classical LST or PSE analysis, since the disturbance paths are not prescribed and thus defining the neutral point becomes problematic. An ill-defined neutral point makes comparing absolute N-factors between different linear methods somewhat arbitrary. Therefore, in this work, we anchor our comparisons of N-factor predictions with experimental measurements of surface pressure for the finned-cone taken from the same quiet tunnel experiments at Purdue described earlier [25].

Full details of the experimental facility, finned-cone wind tunnel model, and quiet tunnel measurements are given in refs. [25] and [26]. The sensor frustum of the finned-cone tunnel model was made to be rotatable about a central shaft. This allowed adjustment of the azimuthal angle of an array of high-frequency PCB pressure transducers mounted flush to the model surface. The main sensor array consisted of 20 PCB sensors aligned at a fixed azimuthal angle along the frustum with equal axial spacing between 0.23 m and 0.38 m from a theoretically sharp nosetip. This main sensor array could be set in 0.5° increments by means of a Vernier scale through a range of 25° − 55° relative to the fin centerline.

An experimental N-factor based on the PCB measurements was calculated from a series of experimental runs, each with the PCB sensor array aligned to a different azimuthal position for the same freestream conditions. For each PCB, the root-mean-squared amplitude of the pressure fluctuations was calculated by integrating the power spectral density (PSD) of the measured pressure fluctuations over a frequency bandwidth of 220-280 kHz and then taking the square root. In this way, the amplitude of the pressure fluctuations at a given frequency, here $f = 250 \pm 30$ kHz, and fixed position on the surface of the model, $(x, \phi)$, can be expressed as

$$p'(x, \phi, f)_{\exp} = \sqrt{\int_{f-\triangle f}^{f+\triangle f} \text{PSD}(x, \phi, f')df'}, \tag{21}$$

where the integral of the PSD is taken over the frequency bandwidth $2\triangle f = 60$ kHz. The magnitude of the measured pressure fluctuations was then normalized by the local maximum at the x-location of the second PCB sensor, i.e., along the wall at $x = 0.24$ m, which is notated as $p'_{\max}$. The natural log of the normalized pressure fluctuations then gives the surface N-factor relative to $p'_{\max}$ and can be expressed as

$$N = ln\left(\frac{|p'|}{|p'_{\max}|}\right) \tag{22}$$

This same procedure of calculating an N-factor relative to the maximum along $x = 0.24$ m was done for all of the stability computations. Figure 13 shows a comparison of measured and computed surface N-factors for a frequency of about 250 kHz, which corresponds to the most unstable vortex mode discussed in section V.A.

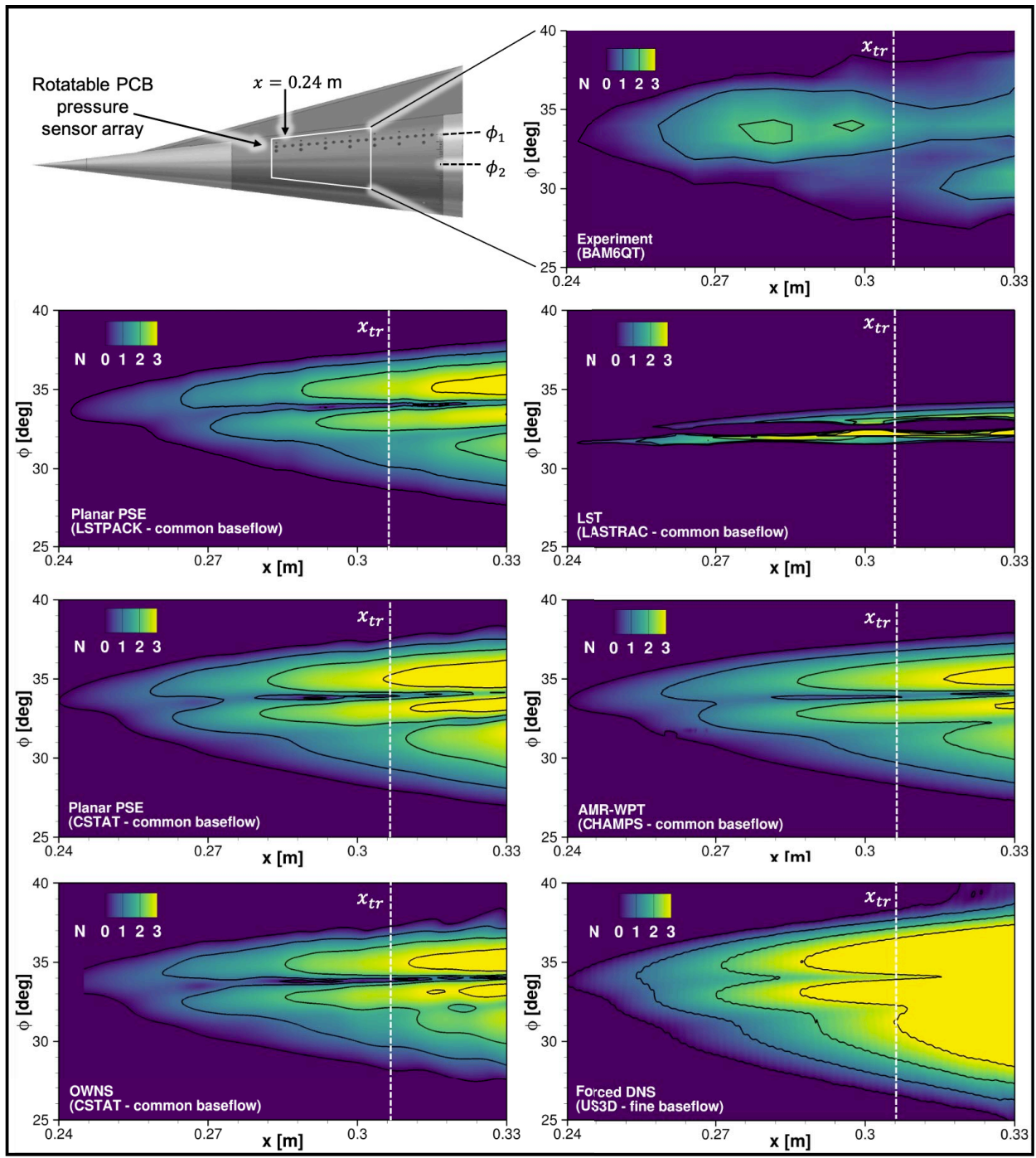

**Fig. 13  Experimental and computational relative N-factors, where N=0 corresponds to the maximum N along $x = 0.24$ m. Experimental N-factor corresponds to measured surface pressure fluctuations at $250 \pm 30$ kHz; data is courtesy of F. D. Turbeville and S. Schneider. Computations of surface N-factor at 250 kHz obtained by LST (LASTRAC), planar PSE (LSTPACK and CSTAT), OWNS (CSTAT), AMR-WPT (CHAMPS), and Forced DNS (US3D) linear stability analysis.**

Qualitatively, there is quite good agreement among the independent planar PSE, OWNS, AMR-WPT, and forced DNS computations shown in figure 13, all of which predict similar features of the surface pressure distribution for the instability, such as the downstream spreading angle of the disturbance and the streamwise-aligned lobes that form. Additionally, the largest predicted N-factors are all aligned close to the $\phi = 35°$ azimuth and correspond to the largest N-factors seen in the experiment. However, there are also some differences worth highlighting, primarily in the range of N-factors that are computed and measured. For the region shown in figure 13, the experimental N-factors saturate at $N \approx 2$, whereas in all of the computations they reach $N \approx 3 - 5$. This difference is attributable to the fact that the flow considered has already transitioned in the experiment, while the computations all remain in the linear regime. Furthermore, compared to the other methods, the LST predictions for this isolated frequency (250 kHz) show a very different surface distribution with almost no (relative) growth predicted outside of a very narrow azimuthal range of $32° < \phi < 34°$; the maximum LST N-factors also appear closer to the fin at $\phi = 32°$, whereas in the experiment and other stability predictions the maximum is slightly more outboard as noted earlier.

Figure 14 shows a more direct comparison of the computed N-factors along the $\phi = 35°$ azimuth. Each of the computations in the figure have been shifted to align with $N = 0$ at $x = 0.24$ m in order to compare with each other as well as the experimental N-factors along this azimuth. Where data was available, we also include N-factor computations upstream of $x = 0.24$ m to illustrate the sensitivity of the predictions to the choice of the neutral point. Note that some computations were initialized at different x-locations and all lines in the figure correspond to what is either stability grid converged or the highest stability grid resolution obtainable for this study, as was described in the methods section.

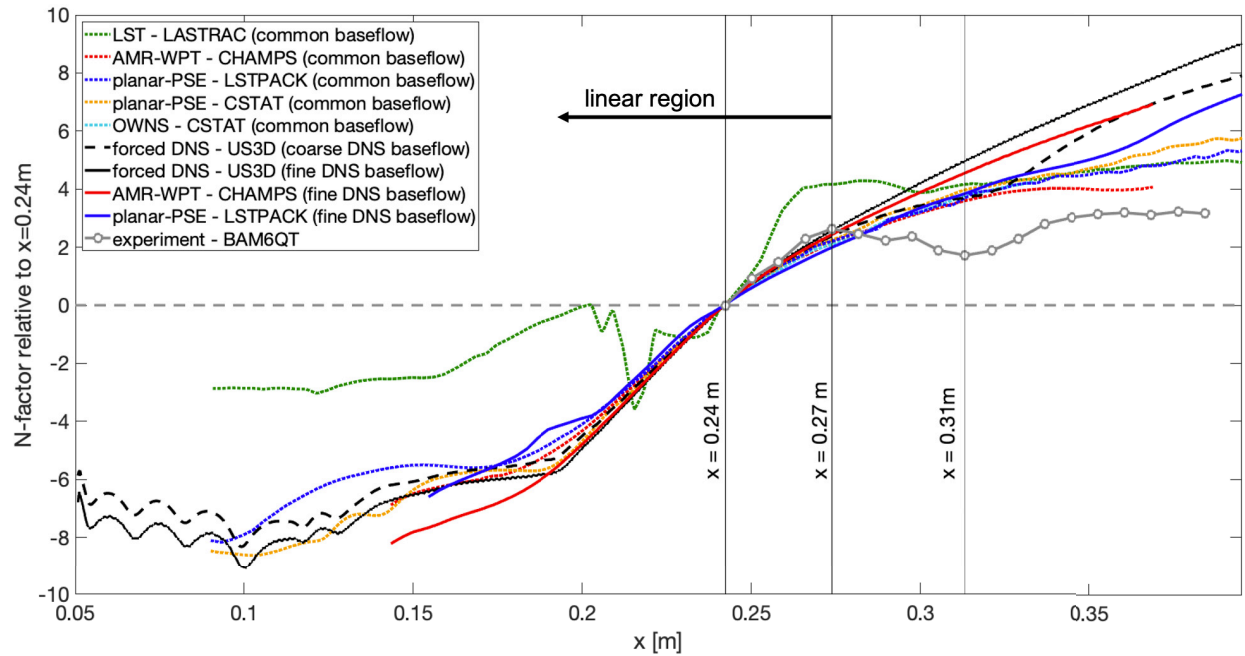

**Fig. 14 Computed and measured N-factor along $\phi = 35°$ relative to $x = 0.24$ m. All curves correspond to a disturbance at a frequency of 250 kHz, with the experimental data including a range of frequencies from 220-280 kHz. Dotted (:), dashed (- -), and solid (–) line types correspond to the common, coarse DNS, and fine DNS baseflow for each stability calculation, respectively.**

29
Distribution Statement A: Approved for Public Release; Distribution Unlimited.

A few observations can be made from figure 14. First, we see that the experimentally measured N-factor reaches a local maximum at $x \approx 0.27$ m, followed by a clear departure from the linear predictions, indicating that the flow has entered a nonlinear stage. Within the measured linear region (i.e., 0.24 m $< x <$ 0.27 m), we observe good agreement between the experiment and each of the linear stability predictions except for LST, which tends to largely over predict the instability growth rate; other linear predictions tend to slightly under predict the growth rate in this region. Between 0.27 m $< x <$ 0.31 m, a significant difference emerges between the forced DNS that utilized the 'fine' baseflow and the other linear predictions that utilized a different baseflow, with larger N-factors predicted downstream by the forced DNS; the LST predictions in this region also follow a different trend with a plateau in the instability growth. Beyond $x > 0.31$ m, there is still good agreement ($\triangle N \lessgtr 1$) among the linear predictions that utilized the same common baseflow. However, the predicted N-factor by the forced DNS with the 'fine' baseflow eventually exceeds the other predictions by as much as $\triangle N = 3$ by the end of the domain. Additional stability computations were carried out utilizing the fine DNS baseflow and are included in figure 14; these tended to show an increase in the predicted N-factor far downstream relative to the predictions that used the common baseflow.

Taken together, the comparisons made thus far strongly support the conclusion that linear predictions of the dominant vortex instability for this flow field are highly sensitive to small changes in the baseflow. For the purpose of this paper in comparing the linear methods to one another, the most important consideration is that stability computations are consistent in utilizing the same baseflow where possible. For more general applications, the results emphasize the critical need to fully resolve all laminar flow features for the most accurate stability predictions.

**C. Mode Shape Comparison**

The mode shape of the pressure fluctuations for the dominant vortex instability is shown for planar PSE, OWNS, AMR-WPT, and forced DNS computations at two different axial stations ($x = 0.25$ m and $x = 0.3$ m) in figure 15. The contours have been normalized by the magnitude of the maximum pressure fluctuation local to each slice and reveal that the disturbance has a unique structure to it with two local peaks, one above the main laminar vortex and another at the wall. There are some slight differences among all of the contours, but generally the figure demonstrates exceptional agreement among the stability computations to predict the mode shape in this region of the flow field. Note that the figure does not include LASTRAC results since the LST analysis is restricted to independent 1D eigenfunction predictions normal to the marching path and so would produce a non-physical 2D mode comparison.

Figure 16 shows a more quantitative comparison of the mode shape among the different methods, where lines at constant $x$ and $\phi$ have been extracted from figure 15 and wall-normal profiles from the LST computations (at at fixed frequency of 250 kHz) have now been added. The most striking feature that can be seen in this figure is the strong peak in the pressure fluctuations that occurs just above the main laminar vortex (at $\eta \approx 2$ mm). This pressure mode shape is distinct from what might be observed for second-mode instability for a sharp cone at zero incidence angle with a peak in the pressure fluctuations typically seen at the wall. For the vortex instability, the peak pressure fluctuations reach approximately 2-3 times the wall pressure and is notably captured by all of the stability methods except LST. These observations again support the earlier suspicion that the locally parallel flow approximation utilized by LST is invalid in this region and a higher fidelity method that can capture strong gradients in both the wall-normal and spanwise directions is required to accurately model the physics.

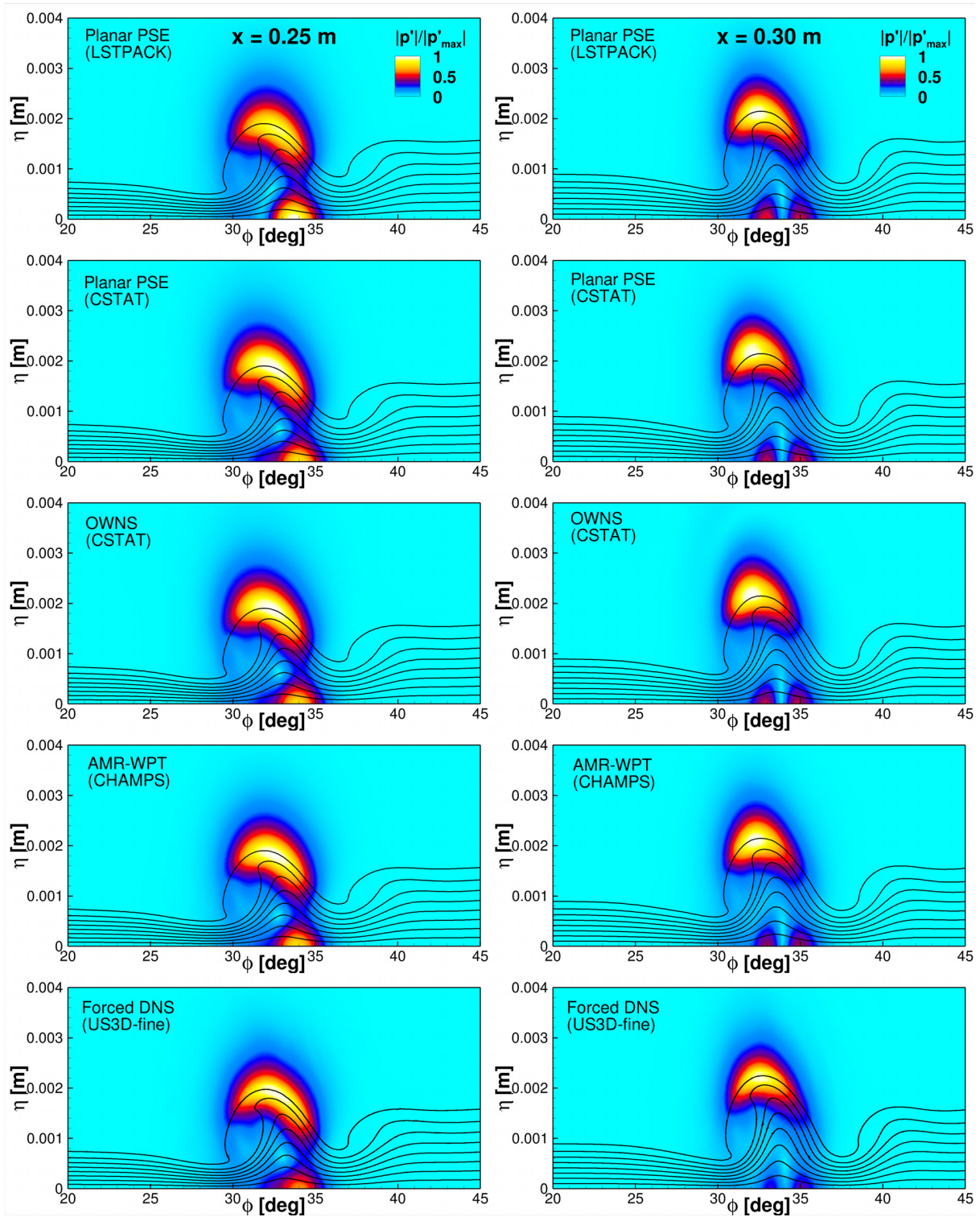

**Fig. 15 Contours of normalized pressure fluctuations calculated by planar PSE, OWNS, AMR-WPT, and forced DNS at a frequency of** $250$ **kHz corresponding to the dominant vortex instability. Mean streamwise velocity contours are overlayed with black lines. All stability computations shown in the figure utilized the common baseflow, except forced DNS, which used the 'fine' DNS baseflow.**

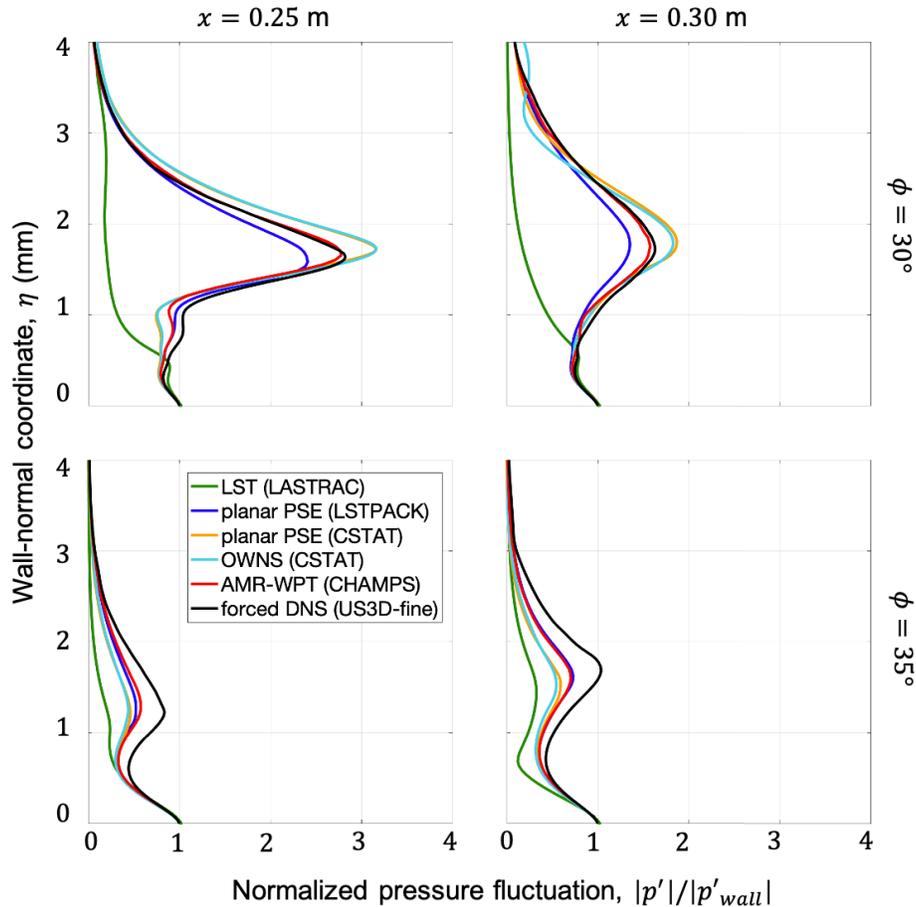

**Fig. 16  Comparison of wall-normal pressure disturbance mode shapes along constant $x$ and $\phi$. All curves correspond to a disturbance at a frequency of 250 kHz and all except forced DNS utilized the common baseflow.**

### D. Effect of Initialization on Instability Development

Due to the added computational cost of global methods, a crucial component to their use is selecting the appropriate type of forcing to use to excite boundary layer instabilities. For this work, both of the global methods that we evaluated used earlier insight from the planar PSE analysis to inform the frequency, structure, and location in the flow field where the instability was likely to grow. Leveraging lower-fidelity stability predictions to guide higher-fidelity analysis is advantageous from a computational cost perspective and is used often in practice. In addition to the planar PSE, we explored the case with OWNS where only the frequency of the dominant mode was known but its spatial development was unknown. One of the major advantages of OWNS over PSE is in its ability to initialize the downstream march with an arbitrary initial condition to allow excitation of all instability mechanisms, including multi-modal interactions and non-modal effects, which is particularly useful for complex flow configurations.

Figure 17 compares N-factors along azimuthal rays of $\phi = 30°$, $32.5°$, and $35°$ between two different OWNS computations. The first OWNS computation uses randomized forcing as an inlet boundary condition to initialize the streamwise march, while the second uses a spatial BiGlobal (SBG) eigenfunction at the inlet; both consider an initial disturbance at a frequency of 250 kHz. Note that $N$ is computed here similarly to Eq. 22, but is normalized with the wall-pressure amplitude at $x = 0.24$ m along the selected azimuthal ray. After the initial transients in the upstream region, both computations converge onto the dominant instability with virtually identical growth rates from $x > 0.2$ m along the three selected azimuths. We also observe that the instability grows the fastest along the $\phi = 35°$ azimuth, resulting in the largest N-factor by the end of the computation.

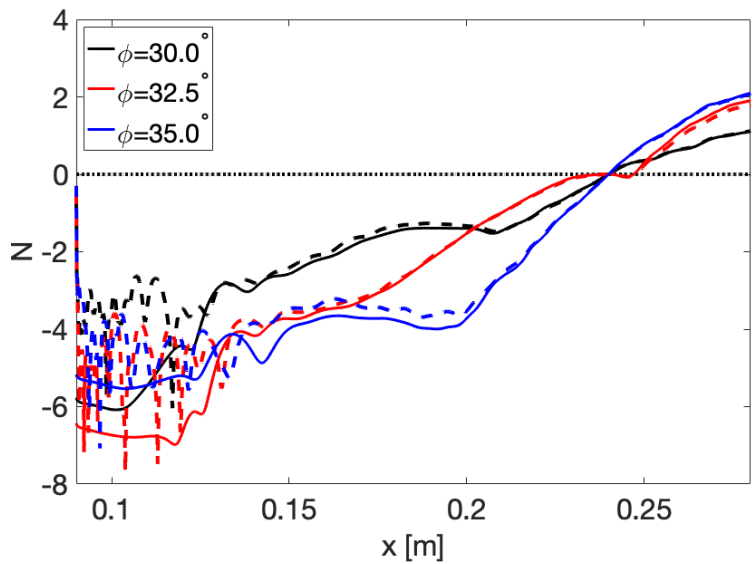

**Fig. 17 Surface N-factors computed using OWNS along prescribed azimuthal rays with randomized inlet forcing (dashed lines) and a SBG inlet boundary condition (solid lines) at $f = 250$ kHz.**

Furthermore, figure 18 shows the shape of the normalized pressure amplitudes for the same OWNS computations shown in figure 17 at three different axial stations ($x = 0.09$ m, $x = 0.185$ m, and $x = 0.28$ m). Although the pressure amplitudes differ vastly at the inlet ($x = 0.09$ m), both calculations indicate the initial development of the vortex instability by $x = 0.185$ m. By $x = 0.28$ m, both computations depict nearly identical pressure amplitude structures corresponding to the growth of the vortex mode identified in the prior sections. We emphasize that the results shown in figures 17 & 18 were obtained using the same method (OWNS) with significantly different initializations (SBG vs. random) and give further confidence to the claim that the dominant instability mechanism for the flow under consideration is indeed attributable to a unique vortex mode.

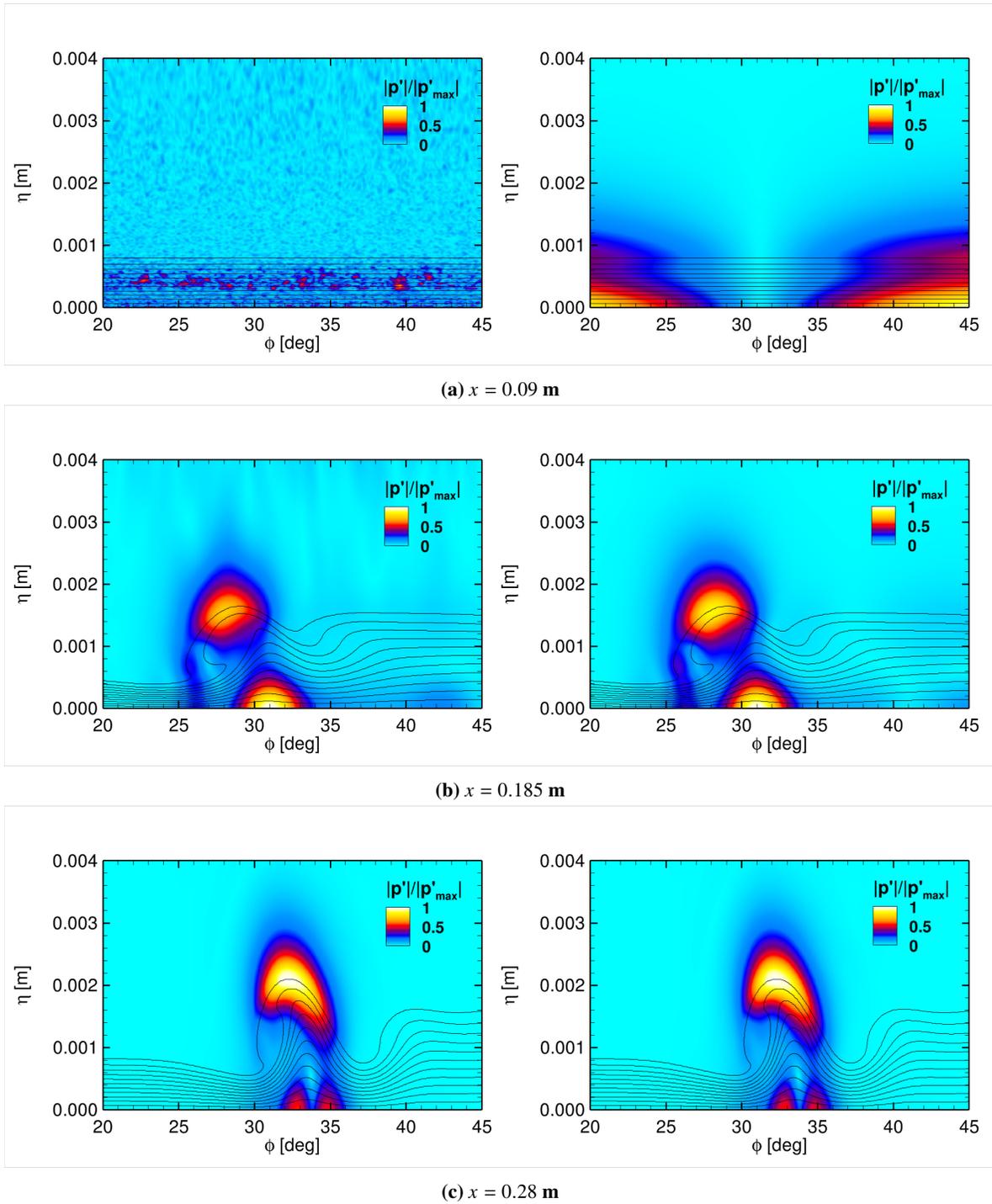

**(a)** $x = 0.09$ **m**

**(b)** $x = 0.185$ **m**

**(c)** $x = 0.28$ **m**

**Fig. 18 Pressure amplitudes computed using OWNS with randomized forcing at the inlet (left) and using a SBG inlet boundary condition (right) at $f = 250$ kHz. The amplitudes are normalized by the maximum value at each streamwise station. The background contour lines correspond to mean streamwise velocity.**

### E. Extension to Nonlinear Breakdown Simulations

Lastly, as was discussed in section II, both AMR-WPT and forced DNS have the potential to extend the simulation of disturbance quantities into the nonlinear regime. For the results presented in earlier sections of this paper, both US3D and CHAMPS were used to solve a set of nonlinear equations with the initial disturbance amplitudes chosen to be small enough to keep the simulations within the linear regime. By increasing this initial disturbance amplitude, the two simulations independently demonstrated signs of nonlinear breakdown occurring, which are depicted in figure 19a for the CHAMPS simulation and 19b for US3D. Although it is not clear how far the simulations have proceeded into the nonlinear regime, a comparison of N-factor curves with the experimental data along $\phi = 35°$ (shown in figures 19a and 19b) suggests that nonlinear saturation of the primary instability mode has set in, which limits the modal growth as compared to the linear simulations (see figure 14). The precise nonlinear breakdown mechanisms that may lead to the over- and undershoots in the wall pressure measurements and experimental N-factor curves is unknown at this stage and would need to be further investigated in a more detailed study that falls beyond the scope of the current paper.

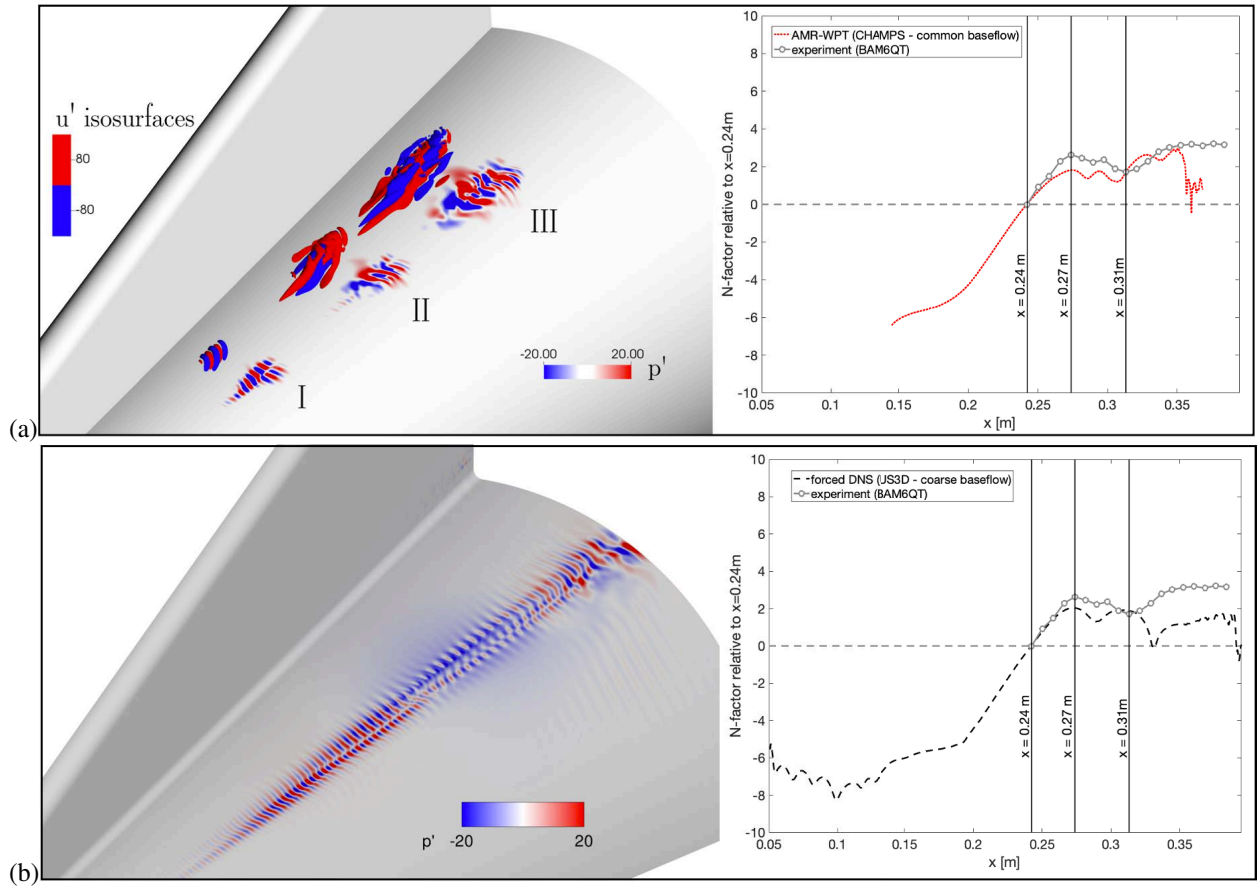

Fig. 19  Possible simulation of nonlinear breakdown (a) by the AMR-WPT (CHAMPS) approach and (b) by the forced DNS (US3D) approach for the finned cone case studied in this paper.

# VI. Concluding Remarks

The goal of this paper is to provide a better understanding of the use of different linear stability analysis methods for analyzing complex hypersonic boundary layer flows. To do so, we analyze a common case of a laminar boundary layer transitioning over the acreage of a finned cone for flow at Mach = 6 and $Re = 8.4 \times 10^6 [m^{-1}]$, as measured in quiet tunnel experiments [25, 26]. Stability results utilizing Linear Stability Theory (LST), planar Parabolized Stability Equations (planar PSE), One-Way Navier Stokes (OWNS), forced direct numerical simulation (DNS), and Adaptive Mesh Refinement Wavepacket Tracking (AMR-WPT) are presented in the paper and the major findings, challenges, and opportunities for future development of these and other methods are discussed in this section.

## A. Major Findings

One of the major findings in this work is the identification of a dominant three-dimensional vortex instability by different stability analysis methods that correlates well with experimental measurements of transition onset. This instability occurs at a frequency of approximately 250 kHz and has a unique mode shape with pressure fluctuations reaching a maximum above the crest of the main laminar vortex formed by the fin-cone interaction region. The peak pressure fluctuations above the vortex for the instability reached approximately 2-3 times the pressure fluctuations at the wall and is distinct from what would be expected for a second-mode instability on a sharp cone at zero incidence angle.

Another major finding is that, apart from classical LST, all of the linear methods tested were consistent in predicting the initial growth and general structure of the vortex instability as it evolved downstream. Interestingly, the LST analysis was able to correctly identify both an unstable frequency range (220-260 kHz) and N-factors as high as 18 in the region of the flow where this vortex instability is dominant (for $\phi = 30 - 35°$). However, the mode shape predicted by LST was inconsistent with the other methods, indicating that the analysis likely gets close to the "correct" answer for the wrong physical reasons near the vortex. LST analysis is by far the most computationally economical and mature approach of the methods considered, so it is reasonable to try as a first attempt in examining boundary layer stability characteristics for a complex flow field. However, when considering design decisions that depend on understanding the transition physics, higher-fidelity methods should be utilized when appropriate.

The results of this study again highlight the critical importance of properly resolving the laminar baseflow for stability analysis. Initially, all of the methods except the forced DNS utilized a common baseflow that was highly resolved in the spanwise direction and produced results that were consistent with each other and with the limited experimental data. The forced DNS approach requires much finer streamwise grid resolution as well as higher-order spatial flux computations and lower dissipative numerics and therefore needed to utilize a different baseflow. This revealed slight differences in resolving the structure of the main laminar vortex that had a significant effect on the amplification of the dominant instability far downstream. Further stability analysis utilizing the fine DNS baseflow revealed larger N-factor predictions and emphasized that the growth of the identified vortex instability is highly sensitive to the baseflow computation.

**B. Challenges and Opportunities**

As was stated in the introduction, it is important to emphasize that the results of this work are for a specific flow field that may be well-suited to one or more of the linear methods or codes utilized and is not intended to be comprehensive or to promote one method or code above the rest. One issue that was not addressed is the computational cost of each of the simulations. Generally speaking, the computational cost goes up with increasing fidelity. However, since each of the methods/codes are at different development stages and have advantages for use depending on the objectives of the analysis, it is very difficult to fairly compare the cost of the computations for the finned cone case that was examined, as it could be misleading and so was not attempted.

Of the lowest fidelity modal methods, namely classical LST and PSE, the main challenge is in knowing where the method can be appropriately applied. Future work aims to develop a type of mean flow metric to characterize a laminar baseflow including its gradient information such that an analyst can quickly assess where an LST or PSE analysis is most appropriate. Such a metric would, however, still rely on starting with a fully resolved baseflow for maximum accuracy. The higher fidelity modal methods, namely BiGlobal and planar PSE, have now been successfully applied to multiple complex geometries and have proven to be very useful. Still, there are several challenges and areas of research that remain. First, a well-known deficiency of ordinary (line-marching) PSE is that the marching step size cannot be reduced indefinitely without incurring numerical instability. This prevents rigorous grid convergence studies because the step size cannot be reduced beyond a certain point. Methods such as pressure gradient suppression have been developed in the case of ordinary PSE to address this [66], but for planar PSE the step size constraints are not known and suitable mitigation techniques have not been developed. Second, planar PSE (like ordinary PSE) can capture only a single mode (with a single streamwise wavelength) at a time. Even for ordinary PSE there are situations where multiple modes exist at the same frequency [36]; for planar PSE such situations are even more common, including the finned cone considered here. Further research is needed to determine whether planar PSE excludes important physics due to modal interactions in these scenarios. Third, the planar PSE approach assumes that instabilities can be described in terms of a single streamwise wavenumber. This is not always possible. When instabilities are present that have a spatially varying streamwise wavenumber, the projection onto a single wavenumber can result in a streamwise-oscillatory shape function, which violates the PSE assumption of a slowly-varying shape function. Methods are needed to screen for this possibility and mitigate it, possibly by judiciously orienting the PSE planes, which changes the component of the total wavenumber that projects onto the streamwise direction.

Moving away from modal methods utilizing OWNS and I/O analysis [22, 41, 68, 69, 79] are promising new approaches to analyzing hypersonic boundary-layer flows that offer several advantages over some of the limitations of planar PSE described above. For one, global I/O analysis has the ability to decompose the global dynamics of an arbitrary three-dimensional flow field into constituent mechanisms that distinguishes it from the other stability approaches. This gives strong motivation to continue to pursue the approach, beyond its natural ability to handle complex geometry with strong flow variations (such as perturbations passing through shocks [79]). The current main challenge for 3D I/O is in identifying efficient algorithms that can solve the sparse linear system with 10s-100s of millions of degrees of freedom, as is common in complex 3D flow fields like for the finned cone. An OWNS-based approach to I/O, i.e., "optimal OWNS", is an alternative to global I/O that has already been successfully demonstrated for hypersonic boundary-layer flows [68, 69]. However, this approach, too, requires significant computational resources and could benefit from more research into efficient algorithms for solving sparse linear systems. The present OWNS computations were performed on a single node due to limitations of the employed PARDISO solver. Future work includes MPI-enabled PARDISO and/or adopting the more efficient solution strategy of [80] to fully exploit the unique and powerful features of the OWNS methodology, including 3D input-output analysis.

Lastly, the highest fidelity methods, namely forced DNS and AMR-WPT, are obviously well-suited for studying complex boundary layer instabilities, but they come at a higher computational cost than the other stability methods discussed, making them currently better suited as supplemental rather than primary tools if required in an engineering design environment. The grid resolution requirements for DNS are typically greater than other methods as the length scales of both the baseflow and boundary layer instabilities must be resolved. The use of targeted grid refinement would help by focusing resolution where needed without wasting grid points elsewhere and has been previously demonstrated for a slightly different finned cone configuration [28]. The use of adaptive grid refinement in the AMR-WPT method clearly reduces the number of grid points required for a high-fidelity transition simulation, but it, too, can be more expensive than some of the lower fidelity transition prediction methods. The use of a pulse disturbance to excite a wide range of frequencies at the same time can reduce the overall cost of high-fidelity simulations. However, the characteristics of the pulse must be chosen carefully in a way to trigger the relevant instability mechanisms. Future development of the AMR-WPT approach aims to involve the adoption of sub-cycling (sometimes referred to as local time stepping), which can provide additional computational savings but are highly dependent on the actual mesh topology. Additional efforts also intend to include the formulation of robust higher-order inter-level AMR operators in combination with higher-order energy stable finite-difference schemes applied to the disturbance flow equations, which are able to conserve certain secondary quantities (such as kinetic energy and entropy to) for additional stability of the numerical implementation. Another challenge with DNS is that it requires time-accurate advancement of the unsteady solution that is comparable in cost to computing the initial steady-state baseflow. Furthermore, a primary consideration when using DNS for investigating boundary layer instabilities is in the interpretation of the data. While linear stability methods like PSE provide a clear means of differentiating instability modes, decomposing DNS data is less straightforward, though modal analysis techniques like POD, DMD, and SPOD [81] can provide insight. To make a meaningful comparison between the results of DNS and linear stability methods, care must be taken to decompose the DNS data in a manner that is consistent with the results of linear stability analysis. Finally, an important consideration for all of the global methods is determining what type of forcing is most appropriate. Leveraging lower-fidelity stability predictions is the current state of practice, with either a trial and error approach using LST or PSE to inform a range of unstable (modal) frequencies and OWNS offering a novel means to introduce arbitrary initial disturbances and track multi-modal and non-modal effects. While this is highly problem dependent, future investigations would benefit greatly from a more unified and rigorous approach to determining how to introduce disturbances into the boundary layer.

## Acknowledgments

The authors wish to acknowledge Drew Turbeville and Steve Schneider for providing the experimental data used in this work. D.B.A, N.P.B, and B.M.W were supported by the Office of Naval Research under NAVSEA contract number N00024-13-D-6400 (PO: Dr. Eric Marineau). O.K. and T.C. were supported by The Boeing Company through the Strategic Research and Development Relationship Agreement CT-BA-GTA-1 and through the Office of Naval Research via grant N00014-21-1-2158. They also acknowledge support of the Natural Sciences and Engineering Research Council of Canada via the Postgraduate Doctoral Scholarship (PGSD3-532522-2019). J.W.N. was supported by ONR grant N00014-19-1-2037. CB and VR gratefully acknowledge funding support provided by ACCESS with grant number 80NSSC21K1117 and Dr. Claudia Meyer as Program Manager and by the Office of Naval Research under contract N00014-19- 1-2223 with Dr. Eric Marineau as Program Manager. A.L.K. and G.V.C. were supported by the Air Force Office of Scientific Research under grant number FA9550-21-1-0106 and the Office of Naval Research under grant number N00014-19-1-2037. The views and conclusions contained herein are those of the authors and should not be interpreted as necessarily representing the official policies or endorsements, either expressed or implied, of the AFOSR, the ONR, or the U.S. Government.